\documentclass[fleqn,usenatbib]{mnras}

\usepackage{newtxtext,newtxmath}

\usepackage[T1]{fontenc}
\usepackage{ae,aecompl}

\usepackage{graphicx}
\usepackage[dvipsnames]{xcolor}
\usepackage{amsmath}
\usepackage{amssymb}
\usepackage{savesym}
\savesymbol{tablenum}
\usepackage{makecell}
\usepackage{siunitx}
\usepackage{xspace}
\usepackage{listings}
\usepackage{soul}
\usepackage{caption}
\restoresymbol{SIX}{tablenum}

\usepackage{comment}

\usepackage{fontawesome}
\definecolor{linkcolor}{HTML}{2AAD2E}

\newcommand\fr{\texttt{frank}\xspace}

\newcommand{\ml}{M$\lambda$}
\newcommand{\kl}{k$\lambda$}
\renewcommand{\vec}[1]{\mathbfit{#1}}
\newcommand{\mat}[1]{\mathbfss{#1}}
\newcommand{\diag}[1]{\mathrm{diag}\,(#1)}
\newcommand{\mean}{\boldsymbol{\mu}\xspace}

\newcommand{\frgithub}{\href{https://github.com/discsim/frank}{\color{linkcolor}{https://github.com/discsim/frank}}}

\title[Frankenstein, the flux reconstructor]{Frankenstein: Protoplanetary disc brightness profile reconstruction at sub-beam resolution with a rapid Gaussian process}

\author[Jennings, Booth, Tazzari, Rosotti, Clarke]{
Jeff Jennings,$^{1}$\thanks{E-mail: jmj51@ast.cam.ac.uk}
Richard A. Booth,$^{1}$ Marco Tazzari,$^{1}$ Giovanni P. Rosotti,$^{2}$ and \newauthor 
Cathie J. Clarke$^{1}$
\\
$^{1}$Institute of Astronomy, University of Cambridge, Madingley Road, Cambridge, CB3 0HA, UK\\
$^{2}$Leiden Observatory, University of Leiden, P.O. Box 9500, Leiden, NL-2300 RA, Netherlands\\
}

\date{Accepted XXX. Received YYY; in original form ZZZ}

\pubyear{2020}

\begin{document}
\label{firstpage}
\pagerange{\pageref{firstpage}--\pageref{lastpage}}
\maketitle

\begin{abstract}
Interferometric observations of the mm dust distribution in protoplanetary discs are now showing a ubiquity of annular gap and ring substructures. Their identification and accurate characterization is critical to probing the physical processes responsible. 
We present Frankenstein (\fr), an open source code that recovers axisymmetric disc structures at sub-beam resolution. By fitting the visibilities directly, the model reconstructs a disc's 1D radial brightness profile nonparametrically using a fast ($\lesssim$1 min) Gaussian process. 
The code avoids limitations of current methods that obtain the radial brightness profile by either extracting it from the disc image via nonlinear deconvolution at the cost of reduced fit resolution, or by assumptions placed on the functional forms of disc structures to fit the visibilities parametrically.
We use mock ALMA observations to quantify the method's intrinsic capability and its performance as a function of baseline-dependent signal-to-noise.
Comparing the technique to profile extraction from a CLEAN image, we motivate how our fits accurately recover disc structures at a sub-beam resolution.
Demonstrating the model's utility in fitting real high and moderate resolution observations, we conclude by proposing applications to address open questions on protoplanetary disc structure and processes.
\href{https://github.com/discsim/frank}{\color{linkcolor}\faGithub} \href{https://discsim.github.io/frank}{\color{linkcolor}\faBook}
\end{abstract}


\section{Physical and methodological contexts}
\label{sec:intro}
High resolution interferometric observations in the sub-mm -- mm and cm are now providing a detailed picture of protoplanetary disc structure, offering an opportunity for new insights into not only disc morphologies but also the underlying physical properties.
Measurements with the Atacama Large Millimeter Array (ALMA) and the Karl G. Jansky Very Large Array (VLA) are finding a ubiquity of annular \lq{}gaps\rq{} and \lq{}rings\rq{} in disc emission. 
At the highest angular resolutions achieved to-date (30 -- 65 mas corresponding to $\approx$5 -- 10 AU)\footnote{Notation: we use $\approx$ to mean \lq{}approximately equal to\rq{} and $\sim$ to mean \lq{}of order.\rq{}}, these features are seen or hinted at in essentially all bright discs around single stars \citep[e.g.,][]{Partnership2015a, Andrews2016, Clarke2018, Huang2018, 2018ApJ...868L...5K, Keppler_2019, 2019NatAs.tmp..419P, 2019arXiv190205143P}. 
Unbiased disc surveys at moderate resolution support this trend, with annular substructure present in at least 38\% (12 / 32) of discs in the Taurus survey at $\approx$120 mas \citep[$\approx$15 AU;][]{Long2018} and hinted at in $\approx$ 5\% (3 / 53) Ophiuchus discs at $\approx$200 mas \citep[$\approx$28 AU;][]{2019MNRAS.482..698C}. Intriguingly this suggests that many apparently smooth discs harbor unresolved gaps and rings, especially compact discs that are covered by a small number of resolution elements in current datasets. Even in the case of the
DSHARP survey at 35 mas \citep[$\approx$5 AU;][]{Andrews2018, Huang2018} there is evidence of only partially resolved substructure.
Together these findings motivate the utility of a high resolution technique that can recover annular features on sub-beam scales and thus inform both theory and follow-up observations.

The ubiquity of annular substructures has prompted a variety of
physical explanations for their origins, invoking forming and newly formed planets \citep[e.g.,][]{1979MNRAS.188..191L, 2012ARA&A..50..211K, 2019NatAs.tmp..329H, Isella_2019} and/or disc processes such as preferential dust growth in localized regions \citep[e.g.,][]{2012A&A...538A.114P, Zhang2015}, opacity effects due to ice sublimation fronts \citep[e.g.,][]{2016ApJ...818L..16Z}, internal photoevaporation \citep[e.g.,][]{Clarke2001, Alexander_2006}, zonal flows \citep[e.g.,][]{Flock2015}, and the vertical shear instability \citep[e.g.,][]{2017ApJ...850..131F}. Highly accurate characterization of axisymmetric features can place constraints on these potential mechanisms for gap and ring creation, including the required dust properties \citep[see applications in, e.g.,][]{2016MNRAS.459.2790R, Fedele2018, Clarke2018, 2018ApJ...869L..46D}.
Even in the presence of non-axisymmetric disc structure such as spirals, accurate recovery of the background radial profile is often an important first step, allowing isolation of the asymmetric features \citep{2016Sci...353.1519P, 2017ApJ...839L..24M}. 
More accurate characterization of azimuthally averaged radial brightness profiles can thus aid in distinguishing the origins of both symmetric and asymmetric morphologies. While this could alternatively be achieved through higher resolution and/or deeper observations, extracting this information from existing datasets is more practically achievable.

This work will present a robust and accurate method to reconstruct an azimuthally averaged radial brightness profile by fitting the interferometric dataset. 
The challenge in this reconstruction
(just as for the 2D brightness) is that the instrument response function resides in the Fourier domain, where the observations are an incomplete (sparse) representation of the source.
Inverse Fourier transforming the visibilities to ultimately obtain a brightness profile thus requires some methodology to infer the unsampled visibilities.
Three standard approaches for modeling protoplanetary disc observations are: use of
the inverse modeling CLEAN nonlinear deconvolution algorithm \citep{1974A&AS...15..417H, 1980A&A....89..377C}; forward modeling regularized maximum likelihood techniques that operate in the image plane, such as the maximum entropy method \citep[e.g.,][]{1986ARA&A..24..127N}; and forward modeling by fitting a model directly to the visibilities. 

Deprojecting and azimuthally averaging a CLEAN 2D model image is the most common technique used to obtain a disc's brightness profile.
CLEAN constructs an empirical model for the sky brightness using a \lq{}dirty image,\rq{} the inverse Fourier transform of the weighted visibilities, which is equivalent to the convolution of the sky brightness with the \lq{}dirty beam\rq{} (the observational point spread function generated from the inverse Fourier transform of the weighted $(u,v)$ distribution).
Since the dirty beam has appreciable sidelobes, this introduces significant artifacts in the dirty image that CLEAN then attempts to remove. This is done by iteratively selecting the brightest pixel in the dirty image and adding a fraction of its amplitude to the model image in the form of a point source or Gaussian \citep{2009GGG....10.9U07A}. This additional source is then convolved with the dirty beam and subtracted from the dirty image. The iteration proceeds until ideally the dirty image contains only noise.
The resulting CLEAN model is then convolved with the \lq{}CLEAN beam\rq{} (typically an elliptical Gaussian fit to the dirty beam's main lobe) to suppress the extrapolation of the model to scales below the beam \citep{1999ASPC..180..151C} and give the final model image (\lq{}CLEAN image\rq{}).
Importantly \emph{this convolution spatially broadens and reduces in amplitude all features in the model image}. The effect is most severe for sub-beam structures but still alters even those resolved by the beam (see Sec.~\ref{sec:convolution}). Beam convolution thus places an intrinsic resolution limit on brightness profile extraction from the disc image. We will use several mock and real datasets to compare profile extraction from a CLEAN image to our model in this work.

An alternative approach to CLEAN is the forward modeling class of regularized maximum likelihood methods (RML). RML techniques for image reconstruction in radio interferometry have been used for decades \citep[e.g.,][]{1974A&AS...15..383A, 1978Natur.272..686G, 1985A&A...143...77C} and have recently evolved rapidly, driven largely by their application to Event Horizon Telescope (EHT) observations and intent to apply them to Next Generation Very Large Array (ngVLA) observations \citep[e.g.,][]{Lu_2014, 10.1093/pasj/psu070, Chael_2016, Johnson_2017, akiyama2019exploring}. The broad approach is to construct an image (among the infinite set of images consistent with the sparsely sampled visibilities) that simultaneously maximizes the likelihood (minimizes $\chi^2$) and minimizes an additional constraint, such as the pixel brightness or image entropy \citep{1986ARA&A..24..127N, 2019ApJ...875L...4E}. Importantly, RML methods have demonstrated success in attaining sub-beam resolution, including in the EHT observations of M87 \citep{2019ApJ...875L...4E}.

A second alternative to CLEAN is to directly fit a forward model in the visibility domain. This avoids the resolution loss from beam convolution in CLEAN by inferring the unconvolved brightness distribution.
Current forward models of this type \citep[e.g., the \texttt{galario} code of][]{2018MNRAS.476.4527T} require the user to specify a parametric functional form for the brightness profile \citep[existing applications include the discs AS~209 and CI Tau;][]{Fedele2017, Guzman2018, Clarke2018}. This approach advantageously allows one to motivate the profile's functional form using a physical model for the disc, such as the dust optical depth and temperature. However when the goal is simply to accurately fit the highest resolution information in a dataset, a parametric approach can face practical limitations.

These limitations are most evident with high resolution (baselines $\gtrsim 5$ {\ml}) data, which almost all show fine structure in their visibilities at long baselines.
The accuracy of the fit to this structure can strongly influence the recovered profile's identification and accurate characterization of sub-beam features.
While there is no intrinsic resolution limit to a parametric fit of the visibilities, choosing a parametric model profile to closely match this long baseline structure is challenging. The functional form is often motivated by the CLEAN image, which for partially resolved, shallow or blended image features is not a trivial choice. 
Moreover the problem's dimensionality is typically high, making an exploration of multiple model profiles expensive. A disc with a single gap and ring has 15 -- 20 free parameters, and more structured discs can require $>$50. Parameter inference with a Markov Chain Monte Carlo (MCMC) sampler in this high dimensional space can take hundreds of CPU hours to converge.
By comparison, a nonparametric approach can offer greater flexibility and speed to accurately fit long baseline, highly structured visibility distributions. We will use the DSHARP observations of AS~209 \citep{Andrews2018} to compare a parametric visibility fit to our nonparametric approach in this work.

To overcome the intrinsic resolution limit of CLEAN and the practical limitations of forward modeling in the visibility domain, we have developed a technique that fits the visibilities directly (avoiding beam convolution) and nonparametrically (affording the flexibility to fit complicated structure in the visibility distribution) to yield a reconstructed brightness profile. This empirical Bayes method, falling between a forward model with a fully explored posterior and a RML approach, imposes no assumptions on the functional form of the disc or its substructures, is autonomous (requiring no iterative, manual tuning of fit parameters), fast, and consistently achieves sub-beam fit resolution.

This work presents \fr, an open source code to reconstruct the 1D radial brightness profile of an axisymmetric protoplanetary disc.\footnote{The code is available at \frgithub\ (and the docs at \href{https://discsim.github.io/frank}{\color{linkcolor}{https://discsim.github.io/frank}}) under the open source GNU Lesser General Public License v3.}
Sec.~\ref{sec:model} details the code's methodology and assesses its prior sensitivities. Sec.~\ref{sec:results} characterizes the model's performance, both intrinsic and as a function of data quality, using mock and real datasets. This includes application to low, moderate and high resolution datasets, and a detailed comparison to brightness profile extraction from a CLEAN image. 
Sec.~\ref{sec:conclusions} concludes by summarizing \fr's properties and outlining use cases for interferometric observations of protoplanetary discs. 

\section{Model}
\label{sec:model}
Our goal is to infer the true brightness $I_\nu(r)$ of a source
under the assumption of azimuthal symmetry.
To do this we will reconstruct $I_\nu(r)$ at a set of radial locations $r_k$ by directly fitting the observed visibilities in the Fourier domain. This is possible by exploiting the properties of the \emph{discrete Hankel transform (DHT)} detailed in Sec.~\ref{sec:fb_dht} and \ref{sec:model_fit_vis}.

We will show, however, that the direct fit is prone to find solutions with strong oscillations on small spatial scales. To solve this we introduce a Gaussian process in Sec.~\ref{sec:GP} to smooth the reconstructed profile. 
Specifically we will assume that the covariance matrix of the Gaussian process can be nonparametrically estimated from the data (visibilities) under the assumption that this matrix is diagonal in Fourier space. The free parameters (diagonal elements) of the matrix can be identified as the power spectrum of the reconstructed brightness profile. A similar approach, but based on log-normal priors, has previously been applied successfully to the more general problem of inferring 2D brightness distributions from radio observations \citep{10.1093/mnras/stt2244, 2015A&A...581A..59J, 2016A&A...586A..76J, 2016arXiv160504317G, 2018arXiv180302174A, 2019A&A...627A.134A}.

We begin by defining the visibility function of a source as sampled by an interferometer as the 2D Fourier transform of the source brightness \citep{1999ASPC..180..151C},
\begin{equation}
    V_\nu(u,v)=\int\int_S I_\nu (l,m)\ {\rm exp}(-2 \pi i (ul + vm))\ {\rm d}l {\rm d}m.
    \label{eqn:vis_fundamental}
\end{equation}
Here $S$ indicates the region of the sky over which the integral is taken (which is assumed to be small, such that $|l^2 + m^2| \ll 1$); and $I_\nu(l,m)$ is the 2D sky brightness at real space antenna coordinates $(l,m)$ and corresponding Fourier domain coordinates $(u,v)$. $(l,m)$ are measured in radians, $(u,v)$ in the observing wavelength $\lambda$.

The necessary first step in fitting the visibilities is deprojection, correcting for the disc's on-sky inclination, position angle (rotation) and phase offset (departure from centering on the origin in the $(u,v)$ plane). For mock observations in this work we will always consider face-on, phase-centered discs. For real observations we will use published geometry values (which we confirm by fitting a 2D Gaussian to the visibilities) to deproject the data prior to applying our model. Importantly, if a disc has an appreciable vertical thickness or if limb darkening from the optically thick surface is important, many of the assumptions underlying a deprojection approach such as fitting a 2D Gaussian to the visibilities would be invalidated.

Then for an azimuthally symmetric function (in our case a deprojected, assumed azimuthally symmetric disc) this 2D Fourier transform between the disc brightness and visibilities reduces to 1D as a Hankel transform with Bessel function kernels
\citep{2000fta..book.....B, 2017isra.book.....T},
\begin{align}
  I_\nu(r) = \int V_\nu(q) J_0(2\upi q r) 2 \upi q\ {\rm d} q, 
  \label{eqn:continuous_I} \\
  V_\nu(q) = \int I_\nu(r) J_0(2\upi q r) 2 \upi r\ {\rm d} r.
  \label{eqn:continuous_V}
\end{align}
Here $r$ is the radial coordinate in the disc, $q = \sqrt{u^2+v^2}$ the baseline distance in the $(u,v)$ plane, and $J_0$ the order 0 Bessel function of the first kind.

\subsection{Representing the brightness profile as a Fourier-Bessel series}
\label{sec:fb_dht}
To evaluate Equations~\ref{eqn:continuous_I} -- \ref{eqn:continuous_V}, we make use of their relation to Fourier-Bessel series via the DHT. For more information about the DHT see \citet{2015JOSAA..32..611B}\footnote{N.B. The definition of the Hankel transform used here differs from that in \citet{2015JOSAA..32..611B} by factors of $2\upi$.}; here we reproduce only the details necessary for our application in Sec.~\ref{sec:model_fit_vis}. 

Imposing the assumption that the real space brightness profile $I_\nu(r) = 0$ beyond some radial distance $R_{\rm out}$, or that the visibilities $V_\nu(q) = 0$ beyond some spatial frequency $Q_{\rm max}$, enables the expansion of $I_\nu(r)$ or $V_\nu(q)$ in a Fourier-Bessel series. Respectively
\begin{align}
I_\nu(r) &= \sum_{k=1}^\infty \alpha_k J_0\left(\frac{j_{0k} r}{R_{\rm out}}\right), \label{eq:fourier-bessel} \\
V_\nu(q) &= \sum_{k=1}^\infty \beta_k J_0\left(\frac{j_{0k} q}{Q_{\rm max}}\right), \label{eq:visib-FB}
\end{align}
where $j_{0k}$ is a scalar representing the $k$th root of $J_0(r)$, i.e., ${J_0(j_{0k}) = 0}$. 

The coefficients $\alpha_k$ in Equation~\ref{eq:fourier-bessel} can be computed via the orthogonality relationship of Bessel functions with subscripts $k$ and $j$,
\begin{equation}
  \int_0^{R_{\rm out}} J_0\left(\frac{j_{0k} r}{R_{\rm out}}\right)J_0\left(\frac{j_{0j} r}{R_{\rm out}}\right) 2 \upi r {\rm d}r = \upi R_{\rm out}^2 J_1^2(j_{0j}) \delta_{jk},
  \label{eq:bessel-orthog}
\end{equation}
where $\delta_{ij}$ is the Kronecker $\delta$ and $J_1$ the first order Bessel function. Substituting Equation~\ref{eq:fourier-bessel} into Equation~\ref{eq:bessel-orthog},
\begin{align}
  \int_0^{R_{\rm out}} I_\nu(r) & J_0\left(\frac{j_{0k} r}{R_{\rm out}}\right) 2 \upi r {\rm d}r \nonumber \\
  &=\sum_{j=1}^\infty \alpha_j \int_0^{R_{\rm out}} J_0\left(\frac{j_{0k} r}{R_{\rm out}}\right)J_0\left(\frac{j_{0j} r}{R_{\rm out}}\right) 2 \upi r {\rm d}r \nonumber \\
  &= \alpha_k \upi R_{\rm out}^2 J_1^2(j_{0k}).
\end{align}
Noting that the left-hand side of this is just the Hankel transform of $I_\nu$, the brightness profile can be written entirely in terms of its visibilities at a specific set of spatial frequencies $q_k = j_{0k} / (2 \upi R_{\rm max})$ and the Fourier-Bessel series coefficients $\alpha_k$. Analogously the visibility at any $q$ can be computed in terms of the brightness at the set of radial locations (collocation points) $r_k = j_{0k}/(2 \upi Q_{\rm max})$ and the Fourier-Bessel series coefficients $\beta_k$,
\begin{align}
  \alpha_k &= \frac{1}{\upi R_{\rm out}^2 J_1^2(j_{0k})} V_\nu\left(\frac{j_{0k}}{2 \upi R_{\rm out}}\right), \\
  \beta_k &= \frac{1}{\upi Q_{\rm max}^2 J_1^2(j_{0k})} I_\nu \left(\frac{j_{0k}}{2 \upi Q_{\rm max}}\right).
\label{eq:beta}
\end{align}

\begin{figure*}
	\includegraphics[width=\textwidth]{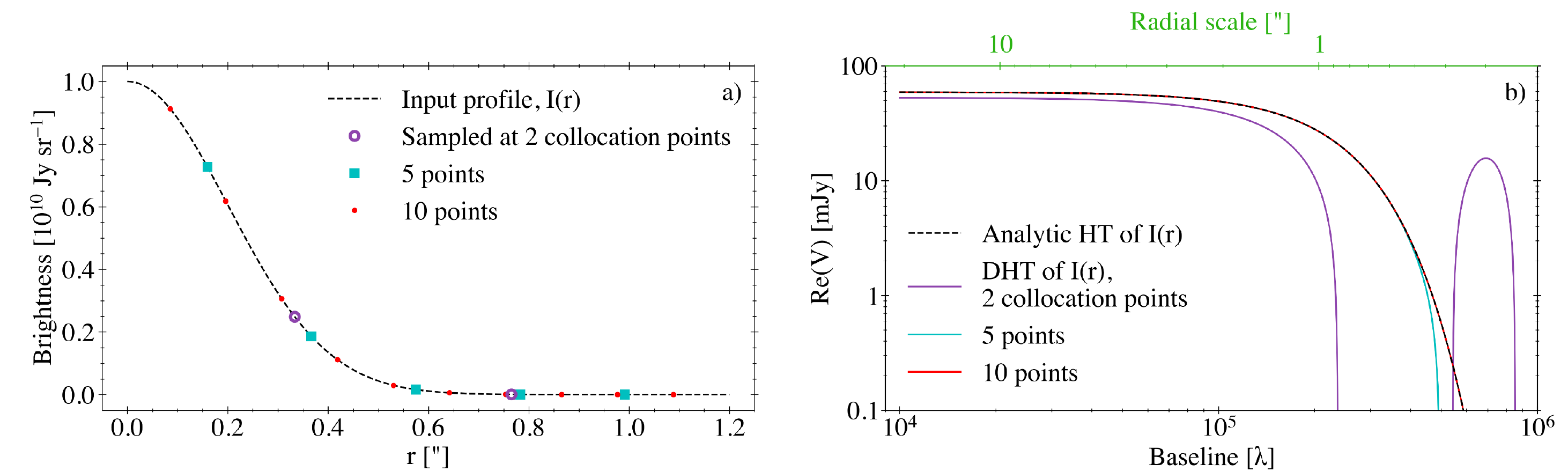}
	    \caption{{\bf Fourier-Bessel series representation of a brightness profile} \newline
	    a) An input mock brightness profile for a Gaussian disc, with samplings at an increasing number of radial collocation points and $R_{\rm out} = 1.2 \arcsec$. \newline
	    b) The discrete Hankel transform (DHT) of a Fourier-Bessel series representation of the input profile at these sets of collocation points, showing that an approximation with a small number of points can closely match the analytic Hankel transform (HT) of the input profile. The second x-axis shows the spatial scale corresponding to a given baseline, $r_{\rm scale} = 1/q$.
        }
    \label{fig:bessel}
\end{figure*}

In practice we must truncate the infinite sums in \autoref{eq:fourier-bessel} and \autoref{eq:visib-FB} to a finite value $N$. From \autoref{eq:fourier-bessel}, the brightness profile then becomes entirely determined by spatial frequencies below some $2 \upi q =j_{0N_+} / R_{\rm out}$, where $j_{0N_+}$ is $j_{0k}$ for $k=N+1$. Similarly the visibilities are entirely determined by radii smaller than $2 \upi r = j_{0N_+} / Q_{\rm max}$.
Choosing $ 2\upi Q_{\rm max} = j_{0N_+} / R_{\rm out}$ then produces the DHT \citep{2015JOSAA..32..611B}. The rules for the backward (visibility space $\rightarrow$ real space) and forward (real space $\rightarrow$ visibility space) transforms of the DHT are 
\begin{align}
  I_k &= \frac{j_{0N_+}}{2\upi R_{\rm out}^2}\sum_{j=1}^N Y_{kj} V_j, \label{eq:Ik} \\
  V_k &= \frac{2\upi R_{\rm out}^2}{j_{0N_+}}\sum_{j=1}^N Y_{kj} I_j, \label{eq:Vk}
\end{align}
where
\begin{align}
  Y_{kj} &= \frac{2}{j_{0N_+} J_1^2(j_{0j})} J_0\left(\frac{j_{0k} j_{0j}}{j_{0N_+}}\right).
\end{align}
The intensities $I_k = I_\nu(r_k)$ and visibilities $V_k = V_\nu(q_k)$ are evaluated at the collocation points of the Fourier-Bessel series in real and visibility space respectively,
\begin{align}
  r_k &= R_{\rm out} j_{0k} / j_{0N_+}, \label{eq:r_coloc} \\
  q_k &= j_{0k} / (2 \upi R_{\rm out}). \label{eq:q_coloc}
\end{align}

We illustrate the correspondence between the brightness profile and visibilities using the example of a Gaussian brightness profile in {\bf Fig.~\ref{fig:bessel}}. A small number of collocation points ($\lesssim 10$, corresponding to the same number of terms in the Bessel series) yields, via the DHT, a visibility profile $V_\nu(q)$ that is in good agreement with the analytic Hankel transform of the input profile up to frequencies $q \sim Q_{\rm max}$. In practice to account for more complicated profiles we use $100 - 300$ collocation points.

It is convenient to absorb the normalization coefficients from Equations~\ref{eq:Ik} -- \ref{eq:Vk} into backward and forward transform matrices respectively\footnote{Notation: We use boldface for matrix quantities, e.g.,
\mat{Y}=$Y_{kj}$.},
\begin{align}
    \mat{Y}_{\rm b} &= \frac{j_{0N_+}}{2\upi R_{\rm out}^2} \mat{Y}, \\
    \mat{Y}_{\rm f} &= \frac{2\upi R_{\rm out}^2}{j_{0N_+}} \mat{Y}, 
\end{align}
which obey
\begin{equation}
  \mat{Y}_{\rm b} \mat{Y}_{\rm f} = \mat{Y} \mat{Y} \approx \mat{I},
\end{equation}
where $\mat{I}$ is the identity matrix. The last approximation is exact only for $N \rightarrow \infty$, though the error is small at modest $N$; for $N > 30$ the largest error is $<$10$^{-7}$. In the code the impact is even less significant because only the forward transform matrices are used explicitly.

These matrices can be used to specify the transformation rules for vectors,
\begin{align}
\vec{f} &= \mat{Y}_{\rm b} \tilde{\vec{f}} , \\
\tilde{\vec{f}} &= \mat{Y}_{\rm f} \vec{f}, 
\end{align}
where we explicitly use a tilde (e.g., $\tilde{\vec{f}}$) to distinguish Fourier domain quantities from a real space vector (e.g., $\vec{f}$). For the visibilities $\vec{V}$ we will drop the tilde.

It will also be useful to define the transformation rules for a covariance matrix $\mat{S}$ and its inverse. These can be derived from the equivalence of scalars in real and visibility space,
$\vec{f}^T \mat{S}^{-1} \vec{f} = \tilde{\vec{f}}^T \tilde{\mat{S}}^{-1} \tilde{\vec{f}}$, where $\tilde{\mat{S}}$ is the visibility space representation of the covariance matrix. From this relation we have
\begin{align}
\tilde{\mat{S}}^{-1} &= \mat{Y}_{\rm b}^T \mat{S}^{-1} \mat{Y}_{\rm b}, \\
\tilde{\mat{S}} &= \mat{Y}_{\rm f}^T \mat{S} \mat{Y}_{\rm f}.
\end{align}

\subsection{Fitting the visibilities using the discrete Hankel transform}

\begin{figure*}
	\includegraphics[width=\textwidth]{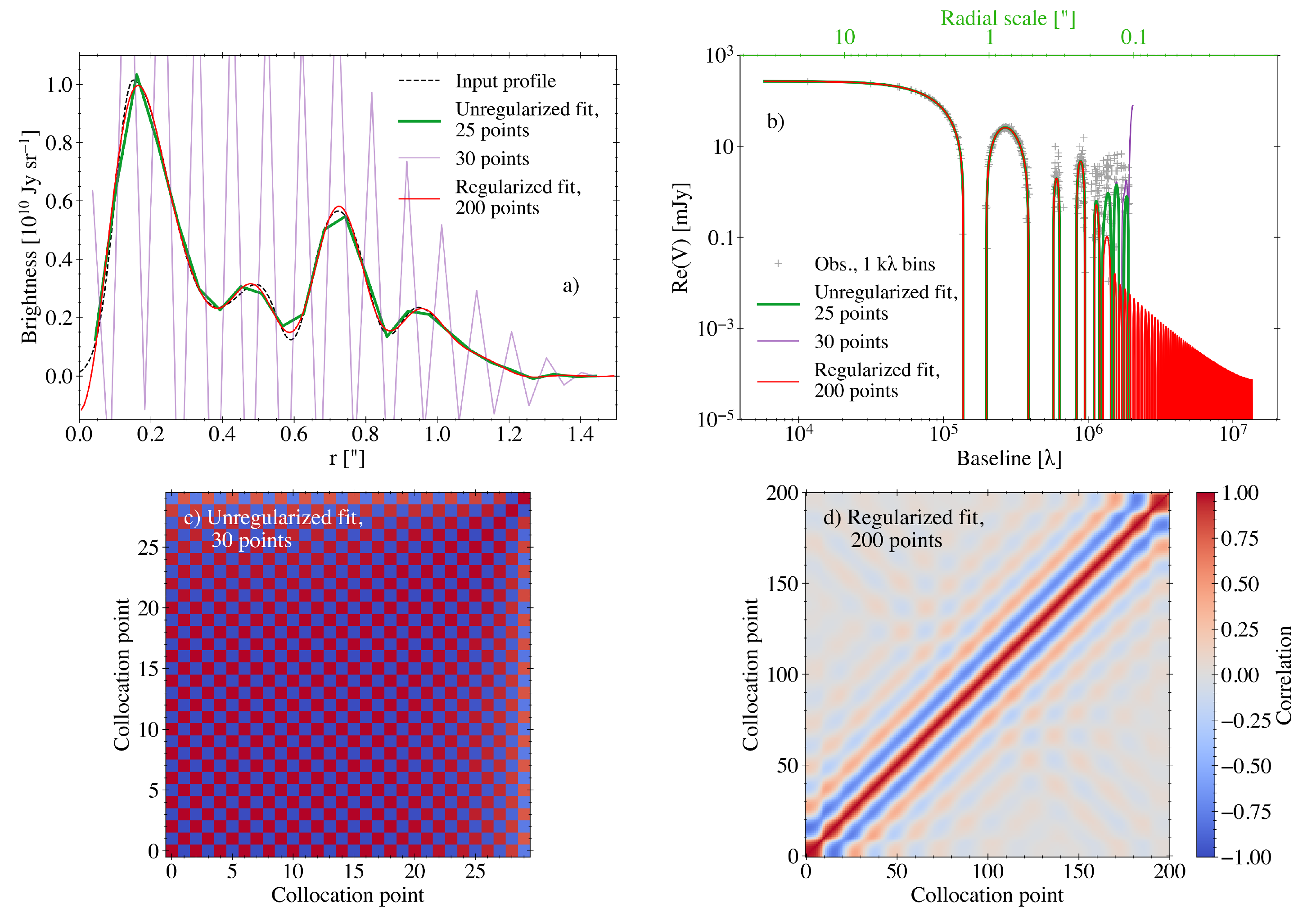}
	    \caption{{\bf Fit regularization} 
	    \newline
	    a) For a multi-Gaussian mock disc observed with the ALMA C43-6 configuration (beam FWHM $0.13 \times 0.17 \arcsec$, Briggs=0.5; see Table~\ref{tab:obs}), a fit with regularization and using 200 collocation points. Fits derived from Equation~\ref{eq:gauss_sol} without regularization are shown for comparison, demonstrating instability at only 30 collocation points. \newline
	    b) Visibilities for the mock observation in (a) and fits corresponding to the brightness profiles in (a). The unregularized fits place erroneously high (noise) power beyond the data's longest baseline, while the regularized fit yields a more reasonable prediction for power on these unobserved scales. \newline
	    c) $30 \times 30$ correlation matrix for the 30-point unregularized fit in (a), showing that the high amplitude oscillations in (b) are a result of almost perfectly strong anticorrelation between adjacent points. \newline
	    d) $200 \times 200$ posterior correlation matrix for the regularized fit in (a), with the regularization providing stability by damping correlations. This in turn prevents regions of erroneously high power in (b) and thus spurious oscillations in (a). 
        }
    \label{fig:regularization}
\end{figure*}

\label{sec:model_fit_vis}
Using the Fourier-Bessel series and DHT from
Sec.~\ref{sec:fb_dht}, we now develop a method to reconstruct a disc's brightness profile given a set of $N_{\rm vis}$ visibilities. To keep the notation succinct we will denote the visibilities by the vector $\vec{V}$, with associated baselines $\vec{q}$ and corresponding weights $\vec{w}$. For a radio interferometer it is reasonable to assume that the noise on each visibility is drawn from an independent Gaussian distribution with variance $\sigma_j^2 = 1/w_j$, such that the covariance matrix of the noise is $\mat{N}= \diag{1/\vec{w}}$. That is, $\mat{N}$ has $1/w_j$ along the diagonal and is 0 otherwise.

Fixing $R_{\rm out}$ (to a distance beyond the edge of the disc) and choosing the number of brightness points $N$ fixes the radial collocation points $r_k$. We can then use the Fourier-Bessel series representation (\autoref{eq:visib-FB}) to link the observed visibilities $\vec{V}$ to $I_\nu(r_k)$, which we seek to infer. The likelihood for $\vec{V}$ is a Gaussian,
\begin{equation}
  \mathcal{L}_{\rm G} = P(\vec{V}|\vec{I}_\nu) =  \mathcal{G}\left(\vec{V} - \mat{H}(\vec{q}) \vec{I}_\nu, \mat{N}\right), \label{eq:like_gauss}
\end{equation}
where generically $\mathcal{G}(\bf{\it{M}}, \boldsymbol{\Sigma})$ refers to a multidimensional Gaussian with mean $\bf{\it{M}}$ and covariance $\boldsymbol{\Sigma}$. The vector $\vec{I}_\nu$ is the brightness at the radial collocation points (i.e., it has the components $I_\nu(r_k)$), and we have introduced the $N_{\rm vis}\times N$ matrix $\mat{H}(\vec{q})$, defined by the components
\begin{equation}
  H_k(q_j) = \frac{4 \upi R_{\rm out}^2}{j_{0N_+}^2 J_1^2(j_{0k})} J_0\left(2 \upi q_j R_{\rm out}\frac{j_{0k}}{j_{0N_+}}\right),
\end{equation}
which comes from the Fourier-Bessel series expansion.

One way to derive $I_\nu(r_k)$ would be to maximize $\mathcal{L}_{\rm G}$. The solution would be
\begin{equation}
  \vec{I}_\nu  = \mat{M}^{-1} \vec{j}\,, \label{eq:gauss_sol}
\end{equation}
where
\begin{align}
  \mat{M} &= \mat{H}(\vec{q})^T \mat{N}^{-1} \mat{H}(\vec{q}),\label{eq:def_M} \\
  \vec{j} &= \mat{H}(\vec{q})^T \mat{N}^{-1} \vec{V}. \label{eq:def_j}
\end{align}
The model's dependence on the visibility data enters entirely through $\mat{M}$ and $\vec{j}$. Note that the construction of $\mat{M}$ scales as $\mathcal{O}(N^2 N_{\rm vis})$, and the construction of $\vec{j}$ scales as $\mathcal{O}(N N_{\rm vis})$, while the solution of these equations and subsequent expressions in Sec.~\ref{sec:GP} scale as $\mathcal{O}(N^3)$, where we recall $N$ is the number of collocation points. Because $\mat{M}$ and $\vec{j}$ are constructed using all unbinned visibilities, the code does not regrid the visibilities onto the spatial frequency collocation points; we are evaluating the full set of observed visibilities. A value of $N \approx 100 - 300 \ll N_{\rm vis}$ is sufficient to fit data at the highest resolutions of current interferometers. 

The problem with maximizing $\mathcal{L}_{\rm G}$ directly is that for sufficiently large $N$, $\mat{M}$ will become singular. This occurs because the relationship between nearby points in the brightness profile is determined by high frequency components. For sufficiently large $N$ these components are either not present in, or are poorly constrained by, the visibilities. Yet the requirement that $N$ is small enough that we are able to invert $\mat{M}$ will be too restrictive, limiting our ability to accurately fit high signal-to-noise (SNR), short baseline data. We would need to fit a large number of data points under the constraint that the profile is smooth on sufficiently small scales. Fitting a brightness profile with a reasonable number of radial points therefore requires regularizing the solution on small spatial scales.

{\bf Fig.~\ref{fig:regularization}} shows how attempting to fit a brightness profile without regularization yields numerical instability in $\mat{M}$ at a modest number of collocation points, $N=30$. While at $N = 25$  the fit is stable, the resulting brightness profile is undersampled, with variations between adjacent radial collocation points. Adding more points to the unregularized fit causes these oscillations to increase in amplitude and frequency, with the visibility domain fit in Fig.~\ref{fig:regularization}(b) having erroneously high amplitude near and beyond the data's longest baselines. In Sec.~\ref{sec:GP} we will describe how the fit can be regularized using a nonparametric Gaussian process model. The regularized fit with \fr shown in Fig.~\ref{fig:regularization}(a) is smooth and insensitive to the number of collocation points (we show the case for $N=200$), yielding an accurate recovery of the input profile. The visibility domain fit in Fig.~\ref{fig:regularization}(b) correspondingly decreases in amplitude at the longest baselines and beyond (note the data are noise-dominated beyond $\approx$1.2 {\ml}).

This difference in behavior between the unregularized and \fr fits is a consequence of the correlations between radial collocation points in each model. In the unregularized fit, adjacent points are almost perfectly anticorrelated for $N = 30$ in Fig.~\ref{fig:regularization}(c),
inducing strong brightness profile oscillations. By contrast the regularized \fr fit introduces a positive correlation
between adjacent points in Fig.~\ref{fig:regularization}(d), damping oscillations in the recovered brightness profile.

\subsection{Regularizing the fit using a nonparametric Gaussian process}
\label{sec:GP}
Regularization corresponds to an a priori assumption that the brightness should be highly correlated at adjacent points and more weakly correlated at distant points. This assumption is well suited to the framework of a Gaussian process, in which the prior on $\vec{I}_\nu$ is a Gaussian,
\begin{equation}
  P(\vec{I}_\nu | \vec{p}) = \mathcal{G}\left(\vec{I}_\nu, \mat{S}(\vec{p})\right), \label{eq:prior_I}
\end{equation}
where $\mat{S}(\vec{p})$ is the prior covariance in the real space brightness profile at the radial collocation points. We explicitly specify that the covariance structure $\mat{S}(\vec{p})$ has some dependence on a set of parameters $\vec{p}$, which we will relate to an estimate of the power spectrum based on the DHT of $\vec{I}_\nu$, $\tilde{\vec{I}}_\nu^2 = (\mat{Y}_{\rm f} \vec{I}_\nu)^2$. Given $\vec{p}$, the posterior probability for $\vec{I}_\nu$ can be used to reconstruct the brightness,
\begin{align}
  P(\vec{I}_\nu|\vec{V}, \vec{p}) &= \frac{P(\vec{V}|\vec{I}_\nu, \vec{p}) P(\vec{I}_\nu|\vec{p})}{P(\vec{V}|\vec{p})}  \label{eq:full_like} \\
  & = \frac{\mathcal{G}\left(\vec{V} - \mat{H}(\vec{q}) \vec{I}_\nu, \mat{N}\right) \mathcal{G}\left(\vec{I}_\nu, \mat{S}(\vec{p})\right)}{P(\vec{V}|\vec{p})}. 
\end{align}
The numerator here is the product of two Gaussians, which is also a Gaussian, and has covariance $\mat{D}$ and mean $\mean$, 
\begin{align}
  \mat{D} &= \left( \mat{M} + \mat{S}(\vec{p})^{-1} \right)^{-1}, \nonumber \\
  \mean &= \mat{D} \,\vec{j}. 
  \label{eq:mean_cond}
\end{align}
Explicitly,
\begin{equation}
P(\vec{I}_\nu|\vec{V}, \vec{p}) \propto \mathcal{G}(\vec{I}_\nu - \mean, \mat{D}),
\end{equation}
which we will use to infer $\vec{I}_\nu$ given $\vec{p}$.
The remaining challenge is how to specify $\mat{S}(\vec{p})$. Typically in a Gaussian process the covariance structure is parameterized in terms of a simple function, such as a Gaussian with some length scale \citep{2006gpml.book.....R}. This length scale could then be optimized or better yet marginalized over. We follow an alternative approach, the empirical Bayes method, in which we use a nonparametric form for the covariance matrix that can be estimated from the visibilities simultaneously with the brightness profile.
This approach follows the work of \citet{2013PhRvE..87c2136O}; see also \citet{2011PhRvD..83j5014E}. 

We make the ansatz that the prior covariance matrix is diagonal in visibility space, i.e., 
\begin{equation}
\tilde{\mat{S}}(\vec{p}) = \mat{Y}_{\rm f}^T \mat{S}(\vec{p}) \mat{Y}_{\rm f} = \diag{\vec{p}},
\end{equation}
thus
\begin{equation}
\mat{S}(\vec{p}) = \mat{Y}_{\rm b}^T \diag{\vec{p}} \mat{Y}_{\rm b},
\label{eqn:Sp}
\end{equation}
where we have now defined the parameters $\vec{p}$ as the diagonal elements of the visibility space representation of the covariance matrix (with the off-diagonal elements set to zero). In the code, we construct $\mat{S}(\vec{p})^{-1}$ directly from $1/\vec{p}$ and $\mat{Y}_{\rm f}$. 

To understand the effect of this prior, we consider $\tilde{\vec{I}}_\nu$ -- the Fourier space representation of $\vec{I}_\nu$ -- which is equivalent to the predicted visibility at the spatial frequency collocation points $q_k$. In visibility space the prior takes the form
\begin{align}
\log P(\vec{I}_\nu | \mat{S}(\vec{p})) &\equiv \log P(\tilde{\vec{I}}_\nu | \tilde{\mat{S}}(\vec{p}))\nonumber \\ 
&= -\frac{1}{2} \vec{I}_\nu^T \mat{S}(\vec{p})^{-1} \vec{I}_\nu - \frac{1}{2} \log{|2 \pi \mat{S}(\vec{p})|} \nonumber \\
&= - \frac{1}{2} \sum_k \left(\frac{\tilde{I}_{\rm \nu,k}^2}{p_k} + \log {p_k}\right)  + {\rm const,} 
\label{eq:log_prior}
\end{align}
where $\tilde{I}_{\rm \nu,k}$ and $p_k$ refer to the $k$th element of $\tilde{\vec{I}_\nu}$ and $\vec{p}$. The last line follows from the definition of $\tilde{\vec{I}}_{\rm \nu}$ and the relation between the determinant of a matrix product and the product of determinants,
\begin{equation}
|2 \upi \mat{S}(\vec{p})| = |\mat{Y}_{\rm b}^T| \cdot |2 \upi \,\diag{\vec{p}}| \cdot |\mat{Y}_{\rm b}| = |\mat{Y}_{\rm b}|^2 \prod_k 2\upi p_k.
\end{equation}

From Equation~\ref{eq:log_prior} we see that if $\tilde{I}_{\rm \nu,k}^2$ (the power in the brightness profile on a scale $k$) is large relative to $p_k$, the prior probability will be small. Thus the prior acts to suppress power on scales where $p_k$ is small. We examine the prior's effect on the reconstructed brightness profile in greater depth in Sec.~\ref{sec:pmap_as_prior}.

\subsection{Jointly inferring the brightness profile and power spectrum parameters}
\label{sec:prior_parameters}
Because we do not know a priori the optimal choice for $\vec{p}$, reconstructing the brightness profile is now a problem of jointly inferring $\vec{I}_\nu$ and $\vec{p}$. The joint posterior probability $P(\vec{I}_\nu, \vec{p} | \vec{V}, \boldsymbol{\beta})$ is constructed using the posterior for $\vec{I}_\nu$ given $\vec{p}$, $P(\vec{I}_\nu | \vec{V}, \vec{p})$, and a prior probability distribution for $\vec{p}$, $P(\vec{p}, \boldsymbol{\beta})$, via 
\begin{equation}
P(\vec{I}_\nu, \vec{p} | \vec{V}, \boldsymbol{\beta}) = P(\vec{I}_\nu | \vec{V}, \vec{p})  P(\vec{p}|\boldsymbol{\beta}).
\label{eqn:joint_prob}
\end{equation}
Here we have noted explicitly the dependence of $P(\vec{p}|\boldsymbol{\beta})$ 
on a set of hyperparameters $\boldsymbol{\beta} = \{\alpha, p_0, w_{\rm smooth}\}$, which also introduces the dependence on $\boldsymbol{\beta}$ into the posterior probability $P(\vec{I}_\nu, \vec{p} | \vec{V}, \boldsymbol{\beta})$. The set of parameters $\boldsymbol{\beta}$ will be held fixed in any given inference of $\vec{I}_\nu$ and $\vec{p}$. We will refer to $P(\vec{p}|\boldsymbol{\beta})$ as the hyperprior to distinguish it from $P(\vec{I}_\nu|\vec{p})$. We define the components of $\boldsymbol{\beta}$ below.

To estimate the parameters of the covariance matrix $\vec{p}$ using the data, our general approach will be to produce small values for $\vec{p}$ on scales that are unconstrained by the data, in order to suppress them, but otherwise allow $\vec{p}$ to be sufficiently large that the reconstructed brightness profile is controlled by the data. 
To achieve this we specify $P(\vec{p}|\boldsymbol{\beta})$ as the product of a spectral smoothness term and inverse Gamma functions,
\begin{equation}
  P(\vec{p} |\boldsymbol{\beta}) = P_{\rm smooth}(\vec{p} | w_{\rm smooth}) \prod_{k=1}^N \frac{1}{p_0 \Gamma(\alpha-1)}\left(\frac{p_k}{p_0}\right)^{-\alpha} \exp\left(-\frac{p_0}{p_k}\right).
  \label{eq:prior_p}
\end{equation}
The exponential part of the inverse $\Gamma$ function disfavors values of $p_k < p_0$, while the power law disfavors $p_k \gg p_0$ for $\alpha > 1$. Neglecting the spectral smoothness hyperprior, the limit $\alpha \rightarrow 1$ and $p_0 \rightarrow 0$ yields a Jeffreys prior (flat in log space). We typically choose a small but nonzero value of $p_0$ (e.g., $10^{-15}\,{\rm Jy}^2$) for practicality; low $p_0$ allows \fr to find a solution with very low power on scales unconstrained by the data, strongly regularizing those scales. Though we do not want $p_0$ to be arbitrarily small, as this leads to numerical instability when evaluating Equation~\ref{eq:mean_cond}.

The spectral smoothness hyperprior follows \citet{2013PhRvE..87c2136O} and is included for two reasons. It first prevents regions of artificially low power arising from narrow gaps in the visibilities and at unconstrained scales beyond the data's longest baseline, ensuring the brightness profile does not exhibit artificially high correlation at the corresponding spatial scales.
Secondly it introduces a coupling between adjacent points in the power spectrum. This has the effect of \lq{}averaging\rq{} the squared visibility amplitude over a range of scales, suppressing the impact of noise on the power spectrum.
Overall we have not found the brightness reconstruction to be highly sensitive to the inclusion of the smoothing hyperprior, which takes the form 
\begin{align}
P_{\rm smooth}(\vec{p}|& w_{\rm smooth}\equiv1/\sigma_{\rm s}^2) \propto \nonumber \\
& \exp \left( -\frac{1}{2\sigma_{\rm s}^2} \int {\rm d} \log(q) \left(\frac{\partial^2 \log(p)}{\partial \log(q)^2}\right)^2  \right).
\label{eq:p_smooth_1}
\end{align}
This hyperprior penalizes power spectra with large second derivatives in log space, i.e., those that deviate from a power law (a straight line in log space). $\sigma_{\rm s}$, which is parameterized in terms of $w_{\rm smooth}$,  controls the allowed amount of variation (departure from a power law) in the power spectrum at a given $q$.
We implement this hyperprior using a numerical estimate of $\partial^2 \log(p)/\partial \log(q)^2$, which can be written in the form
\begin{equation}
P_{\rm smooth}(\vec{p}| w_{\rm smooth}) \propto \exp \left( -\frac{1}{2}\log(\vec{p})^T \frac{\mat{T}}{\sigma_{\rm s}^2} \log{(\vec{p})}\right),
\label{eq:psmooth}
\end{equation}
where $\mat{T}$ is a constant, pentadiagonal matrix that depends only on the spatial frequency collocation points $q_k$. For the exact form of $\mat{T}$, see Appendix~\ref{Sec:T_matrix}.

With $P_{\rm smooth}(\vec{p}| w_{\rm smooth})$ specified, we now have a form for $P(\vec{p}|\boldsymbol{\beta})$. 
Ideally we would proceed to obtain the posterior for $\vec{I}_\nu$ by marginalizing over $\vec{p}$, but the high dimensionality of the parameter space makes this impractical. We instead maximize $P(\vec{p} | \vec{V}, \boldsymbol{\beta})$ to obtain and use the maximum a posteriori value $\vec{p}_{\rm MAP}$ as the prior on $\vec{I}_\nu$. The marginal posterior $P(\vec{p} | \vec{V}, \boldsymbol{\beta})$ is obtained from $P(\vec{I}_\nu, \vec{p} | \vec{V}, \boldsymbol{\beta})$ by integrating over all $\vec{I}_{\nu}$, i.e., $\int P(\vec{I}_\nu, \vec{p} | \vec{V}, \boldsymbol{\beta})\ {\rm d}\vec{I}_\nu$.
Since $P(\vec{I}_\nu|\vec{V}, \vec{p})$ is a multivariate Gaussian, this can be done analytically \citep[e.g., Appendix A of][]{2006gpml.book.....R} as in \citet{2013PhRvE..87c2136O},
\begin{align}
  \log P(\vec{p} | \vec{V}, \boldsymbol{\beta}) =& \frac{1}{2}\,\vec{j}^T \mat{D}\, \vec{j} + \frac{1}{2} \log|\mat{D}| - \frac{1}{2} \log |\mat{S}(\vec{p})| \nonumber \\
  & - \sum_k\left[(\alpha - 1) \log p_k + \frac{p_0}{p_k} \right] - \frac{1}{2} \log (\vec{p})^T \frac{\mat{T}}{\sigma_{\rm s}^2}\log (\vec{p}) \nonumber \\
  &+ {\rm const.} \label{eq:like_power}
\end{align}
\citep[see also][]{2011PhRvD..83j5014E}. Finding the maximum entails finding the location where the derivative of \autoref{eq:like_power} with respect to $\log p_k$,
\begin{align}
  \frac{{\rm d} \log P(\vec{p} | \vec{V}, \boldsymbol{\beta})}{{\rm d} \log p_k} = 
  & \frac{1}{2 p_k}\left[ \left(\mat{Y}_{\rm f} (\mean \mean^T + \mat{D}) \mat{Y}_{\rm f}^T\right)_{kk}\right] - \frac{1}{2}
  \nonumber \\
  & - \left[(\alpha - 1) - \frac{p_0}{p_k}\right] -  \left(\frac{\mat{T}}{\sigma_{\rm s}^2} \log \vec{p}\right)_k,
\end{align}
is zero. 
We find the maximum using the fixed point iteration
\begin{align}
  (\mat{I} +& \frac{\mat{T}}{\sigma_{\rm s}^2})\log\vec{p}^{\rm new} = \log{\vec{p}} \nonumber \\
  & + \frac{1}{p_i}\left[p_0 + \frac{1}{2} {\rm diag} \left(\mat{Y}_{\rm f} (\mean \mean^T + \mat{D}) \mat{Y}_{\rm f}^T\right) \right]
  - \left[(\alpha - 1) + \frac{1}{2}\right] \label{eq:iter}
\end{align}
(recall that $\mat{I}$ is the identity matrix). 
Here the application of the $\diag{\mat{M}}$ operator to a matrix should be understood as selecting the vector formed from the diagonal elements of that matrix. At each iteration $\mean$ and $\mat{D}$ are computed using $\vec{p}$ from the previous iteration, and the linear system is solved using a sparse linear solver. Each iteration requires $\mathcal{O}(N^3)$ operations. The iterations are terminated when the relative change in $\vec{p}$ is $<10^{-3}$. We have confirmed with tests that the final solution is not sensitive to the initial choice of $\vec{p}$; we use as an initialization a power law with slope of $-2$ to coarsely match the typical decline in visibility amplitude as a function of baseline in high resolution ALMA observations.

\begin{figure}
	\includegraphics[width=\columnwidth]{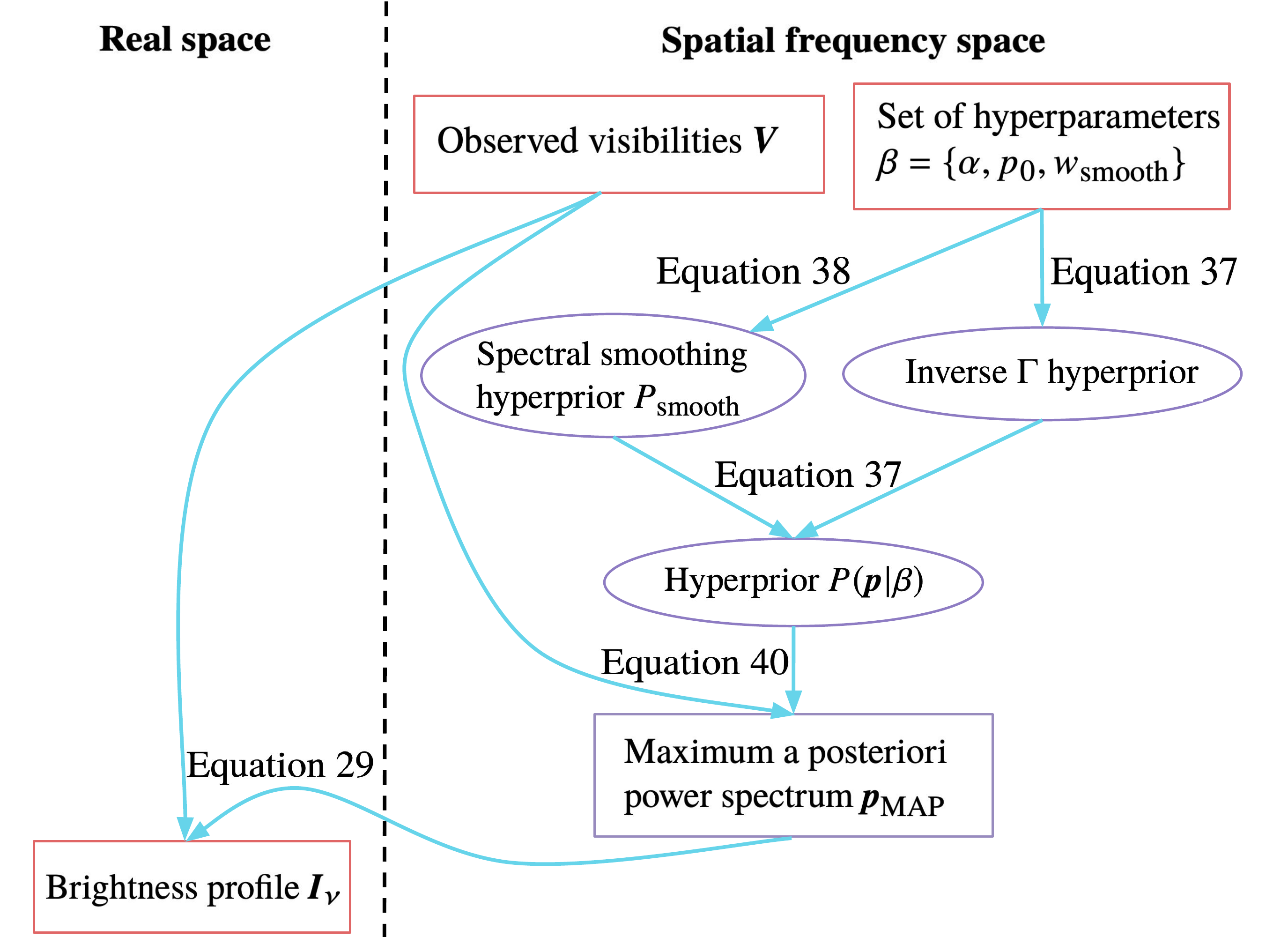}
	    \caption{{\bf Model framework} 
	    \newline
        Diagram of the probabilistic model framework in \fr. Squares represent variables, purple indicates the quantity is either a hyperprior or a prior (note $\vec{p}_{\rm MAP}$ is an inferred variable that is used as a prior). The set of hyperparameters $\boldsymbol{\beta}$ determine the hyperpriors placed on the power spectrum reconstruction from $\vec{V}$, yielding $\vec{p}_{\rm MAP}$. This is then used as a prior for the reconstruction of $I_\nu$ from $\vec{V}$.
        }
    \label{fig:model_diagram}
\end{figure}

For the final reconstructed brightness profile, we use our best-fitting (maximum a posteriori) values for the power spectrum coefficients $\vec{p}_{\rm MAP}$ in \autoref{eq:mean_cond}. This provides our estimate for the profile's mean $\mean$ and covariance $\mat{D}$. 
In general $\diag{\mat{D}}$ will underestimate the uncertainty on the brightness at each collocation point, as discussed in Sec~\ref{sec:model_limitations}. We summarize the overall model framework diagrammatically in {\bf Fig.~\ref{fig:model_diagram}}.

\begin{figure*}
	\includegraphics[width=\textwidth]{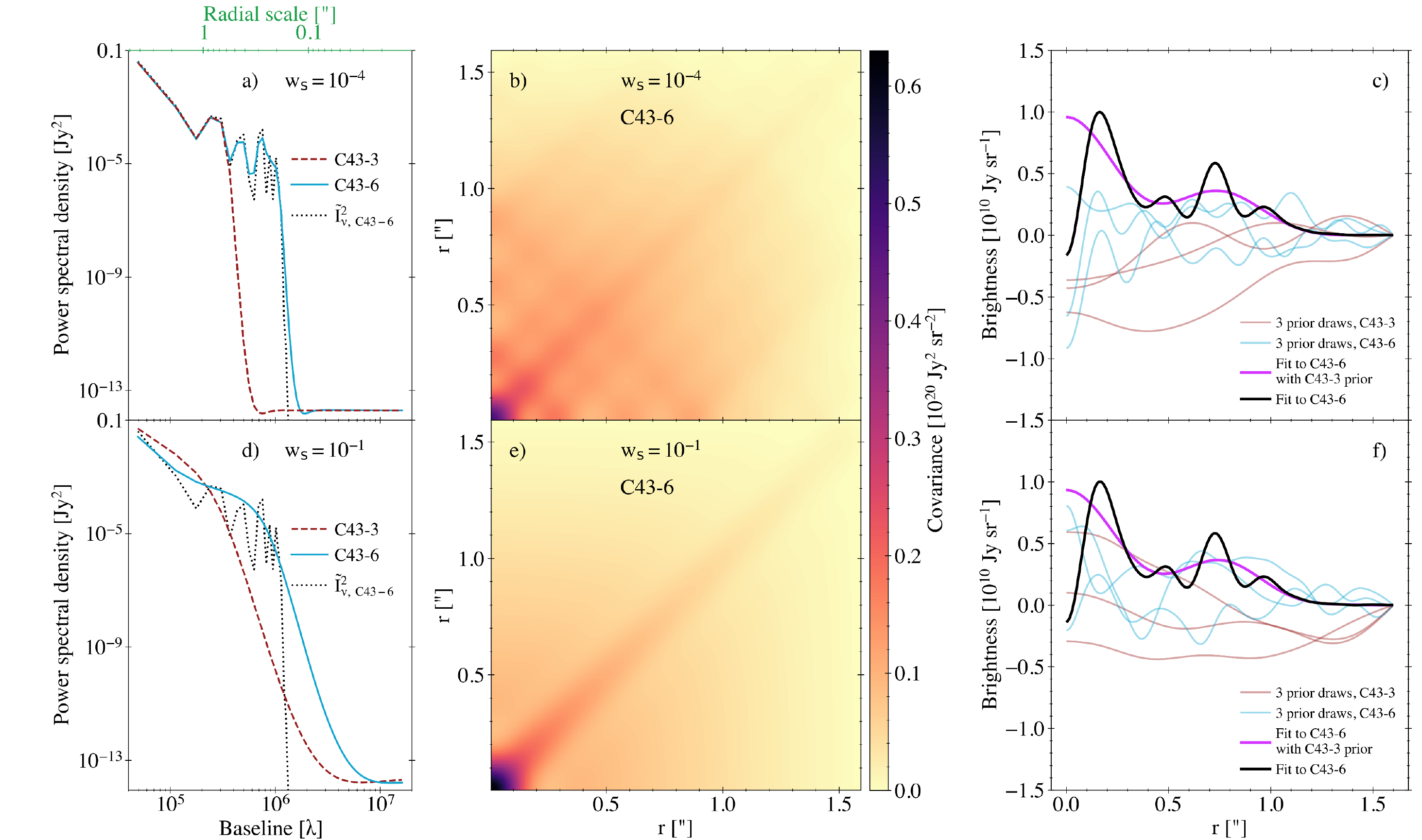}
	    \caption{{\bf Effect of the hyperparameter $w_{\rm smooth}$} 
	    \newline
	    a) Maximum a posteriori power spectra $\vec{p}_{\rm MAP}$ for a multi-Gaussian mock disc observed separately with the ALMA C43-3 (synthesized beam FWHM $0.59 \times 0.70 \arcsec$, Briggs=0.5) and C43-6 (beam FWHM $0.13 \times 0.17 \arcsec$, Briggs=0.5) configurations (see Table~\ref{tab:obs}). These $\vec{p}_{\rm MAP}$ are obtained for each dataset using $P(\vec{p}| \vec{V}, \boldsymbol{\beta})$, the \fr posterior marginalized over all realizations of $\vec{I}_\nu$, under the hyperparameters $\boldsymbol{\beta}=\{\alpha, p_0, w_{\rm smooth}\} = \{1.05, 10^{-15}\,{\rm Jy}^2, 10^{-4}$\}. For comparison the power spectrum estimate $\tilde{\vec{I}}_{\nu,\ {\rm C}43-6}^2 = (\mat{Y}_{\rm f} \vec{I}_{\nu,\ {\rm C}43-6})^2$ based on the \fr fitted brightness profile for the C43-6 dataset is shown. \newline
	    b) Real space representation of the covariance matrix $\mat{S}(\vec{p}_{\rm MAP})$ for the C43-6 configuration (that for the C43-3 looks qualitatively similar), showing covariance between not only adjacent but also nonadjacent points. The overall covariance decreases from the lower left to upper right because the disc brightness as represented in the visibilities decreases (on average) as a function of disc radius. \newline
	    c) Draws from the prior on the brightness distribution $\mathcal{G}(\vec{I}_\nu, \mat{S}(\vec{p}_{\rm MAP}))$ for both the C43-3 and C43-6 datasets. For reference the posterior mean (the \fr fit for the brightness profile, Equation~\ref{eq:mean_cond}) is shown for the C43-6 dataset, separately realized using $\vec{p}_{\rm MAP}$ estimated from the C43-6 and C43-3 dataset. \newline
	    d) -- f) As in (a) -- (c) but using $w_{\rm smooth} = 10^{-1}$ to generate the power spectra, which are comparatively smooth in response to this stronger hyperparameter value, resulting in reduced covariance between nonadjacent points but higher covariance between adjacent points in $\mat{S}(\vec{p}_{\rm MAP})$ as visualized in (e).
	    The reduced covariance between nonadjacent points in (e) relative to (b) results in draws from the prior in (f) that show fewer small amplitude amplitude oscillations on corresponding spatial scales. However the effect on the posterior mean \fr fit is small, indicating the relative insensitivity of the recovered brightness profile to the value of the hyperparameter $w_{\rm smooth}$.
        }
    \label{fig:prior_wsmooth}
\end{figure*}

\subsection{$\vec{p}_{\rm MAP}$ and its use as a prior on $\vec{I}_\nu$}
\label{sec:pmap_as_prior}
The maximum a posteriori power spectrum $\vec{p}_{\rm MAP}$ tends to one of two limiting forms, depending primarily on the visibility SNR at each spatial frequency. To understand this we consider the converged maximum a posteriori values (i.e., $\log p^{\rm new} = \log p$), neglecting the spectral smoothness hyperprior for now ($w_{\rm smooth} \rightarrow 0$).

First, at spatial frequencies where the visibility SNR is sufficiently high,
the term involving the mean $\mean$ in Equation~\ref{eq:iter} dominates and thus
\begin{equation}
  \vec{p}_{\rm MAP} = \frac{(\mat{Y}_{\rm f} \mean)^2}{1 + 2 (\alpha-1)} \label{eq:sol_ps1}.
\end{equation}
In this case the power spectrum is set by an estimate using the DHT of the mean brightness profile,
which is approximately the square of the visibility amplitude (its power). This demonstrates the aforementioned association between $\vec{p}_{\rm MAP}$ and the power spectrum and provides the justification for calling $\vec{p}_{\rm MAP}$ the power spectrum.

Conversely, where the visibility SNR is sufficiently low, $\vec{p}_{\rm MAP}$ is small enough that
the $\mat{S}(\vec{p})$ term in $\mat{D}$ dominates. In this case
\begin{equation}
\vec{p}_{\rm MAP} = \frac{p_0 + \frac{1}{2}\diag{\mat{Y}_{\rm f} \mat{S}(\vec{p}) \mat{Y}_{\rm f}}}{\alpha-1 + 1/2} = \frac{p_0 + \vec{p}_{\rm MAP}/2}{\alpha-1 + 1/2} = \frac{p_0}{\alpha-1}.  \label{eq:sol_ps2}
\end{equation}

The SNR threshold that separates these behaviors depends on $\alpha$ (Appendix~\ref{sec:SNR_thresh}); at $\alpha=1$ the threshold is at SNR $ = 1$, and the threshold increases as $\alpha$ increases (because for larger $\alpha$, the inverse $\Gamma$ hyperprior decays faster with increasing $p_k$). Thus given the same data, a larger $\alpha$ will cause $p_k$ to be pulled more strongly downward toward $p_0$.

\begin{figure}
	\includegraphics[width=\columnwidth]{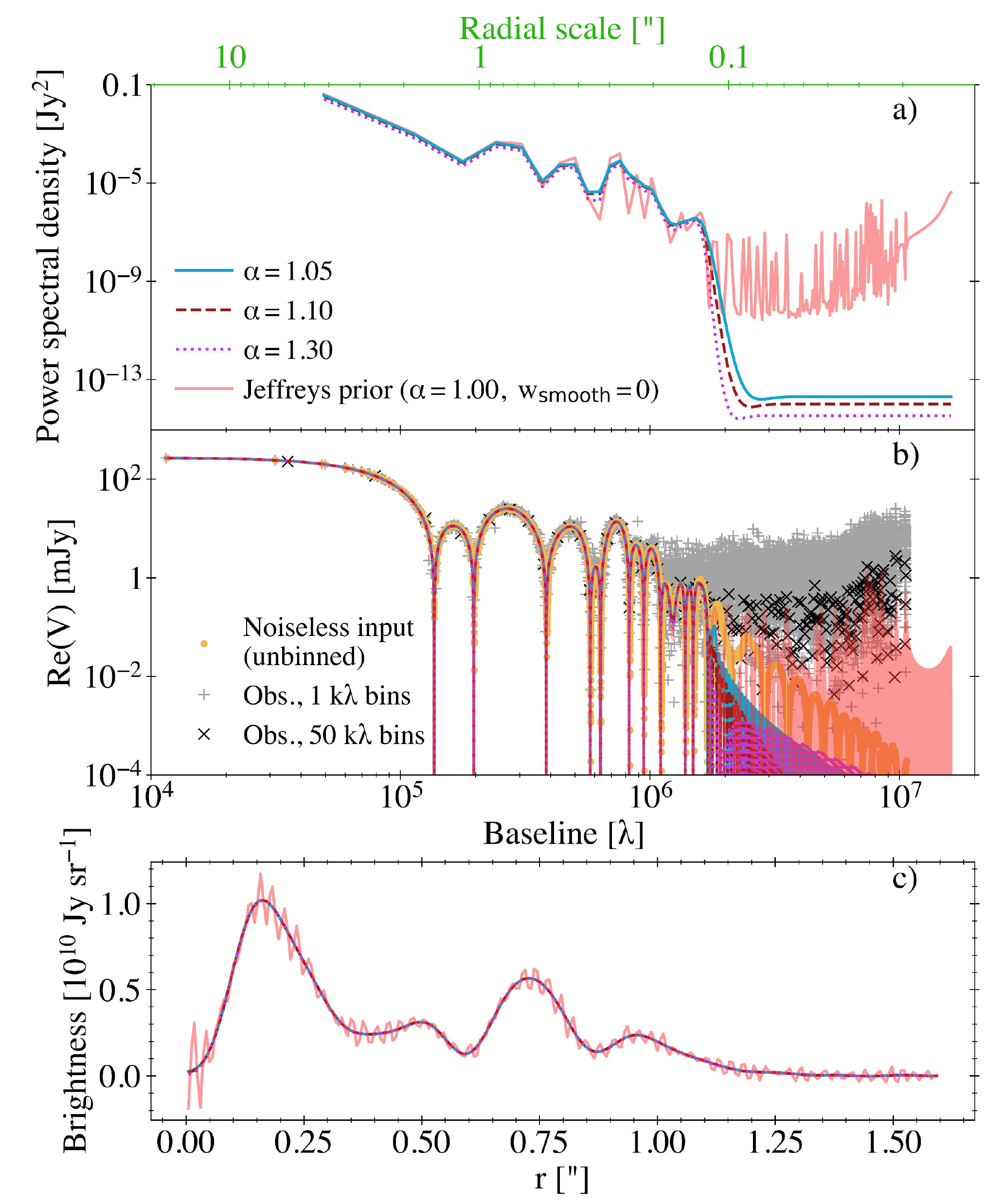}
	    \caption{{\bf Effect of the hyperparameter $\alpha$; effect of a Jeffreys prior}
	    \newline
        a) Maximum a posteriori power spectra $\vec{p}_{\rm MAP}$ (under different values of the $\alpha$ hyperparameter) for \fr fits to the mock disc in Fig.~\ref{fig:prior_wsmooth}, here observed with the ALMA C43-6 + C43-9 configurations (beam FWHM $0.024 \times 0.030 \arcsec$, Briggs=0.5; see Table~\ref{tab:obs}). Also shown is $\vec{p}_{\rm MAP}$ under a Jeffreys prior. \newline
	    b) The mock observation's noisy visibilities in 1 and 50 k$\lambda$ bins and the true, unbinned visibility distribution (the Hankel transform of the input profile). Also shown are the \fr fits to the unbinned, noisy data and the fit under a Jeffreys prior (which forces the fit to the longest baseline, noise-dominated visibilities, imprinting oscillations on the corresponding brightness profile in (c)). \newline
	    c) The fitted \fr brightness profiles for the cases in (a) -- (b). Varying $\alpha$ has a negligible effect on the fit, while the Jeffreys prior (flat in log space) allows oscillations on spatial scales corresponding to the noise-dominated visibility region (beyond $\approx$1.5 {\ml}).
        }
    \label{fig:prior_alpha}
\end{figure}

Secondary to the effect of the visibility SNR, the spectral smoothness hyperprior modifies the power spectrum. Fig.~\ref{fig:prior_wsmooth}(a) and (d) (presented below) demonstrate that increasing $w_{\rm smooth}$ reduces structure in the power spectrum, driving its shape toward a power law. 
A similar effect is evident in Fig.~\ref{fig:prior_alpha} (presented in Sec.~\ref{sec:param_sensitivity}), where the oscillations in the power spectrum under a Jeffrey's prior (using $w_{\rm smooth} = 0$) are large compared with those generated under $w_{\rm smooth} = 10^{-4}$. This demonstrates that in regions of low SNR at long baseline, the constraint of a smooth power spectrum can dominate the inverse $\Gamma$ hyperprior's preference for low $p_i$. 
Though this is typically isolated to the case of $\alpha=1$, because the inverse $\Gamma$ hyperprior does not damp the power spectrum coefficients under this choice.

Given a form for the maximum a posteriori power spectrum $\vec{p}_{\rm MAP}$, we now consider its effects as a prior on the reconstructed brightness profile $I_\nu$.
{\bf Fig.~\ref{fig:prior_wsmooth}} shows multiple power spectra generated using \fr fits to the mock brightness profile in Fig.~\ref{fig:regularization}. Under different ALMA configurations and values of $w_{\rm smooth}$, the power spectra in Fig.~\ref{fig:prior_wsmooth}(a) and (d) are truncated (drop off) at different maximum baselines (and thus minimum spatial scales) and also show different degrees of smoothness.
The covariance matrices $\mat{S}(\vec{p})$ in Fig.~\ref{fig:prior_wsmooth}(b) and (e), and the draws from the prior $P(\vec{I}_\nu|\vec{p})$ in Fig.~\ref{fig:prior_wsmooth}(c) and (f), then motivate the two effects that $\vec{p}_{\rm MAP}$ has as a prior on $I_\nu$.

First, because the power spectra have little power on long baselines, the prior draws are correlated on small spatial scales. The shortest length scale over which a brightness profile is correlated is controlled by the longest baseline at which the prior has significant power. In contrast, because the amplitude of the power spectra is large on short baselines, the large scale form of the brightness is free to vary, with the posterior brightness on these scales being ultimately determined by the data rather than the prior. If instead adjacent points were not correlated (if the prior were diagonal in real space), the prior draws would not appear smooth. Fig.~\ref{fig:prior_wsmooth}(c) and (f) additionally show the posterior means (Equation~\ref{eq:mean_cond}) -- the fitted brightness profiles for the C43-6 mock data using either the C43-3 or C43-6 prior -- which reflect constraints introduced by the priors. For each of the priors, the mean matches the true (input) brightness on large scales, but only shows structure on the scales allowed by the prior. For the priors generated under the C43-3 mock observations, the power on long baselines is thus too strongly damped, and the reconstructed profiles are a poor recovery of the input profile. By contrast the reconstructed profiles using the C43-6 priors recover the input profile to high accuracy. This emphasizes the importance of correctly identifying the scale on which to regularize the brightness profile so as not to damp the fit on scales where true variations exist in the visibility distribution. 

Second, substructure in the power spectrum causes differences in the real space representation of the prior. The localized areas of lower power in the structured power spectra of Fig.~\ref{fig:prior_wsmooth}(a) introduce correlations between nonadjacent radial collocation points at the corresponding spatial scales. This correlation manifests as the crosshatching in Fig.~\ref{fig:prior_wsmooth}(b), and consequently draws of $\vec{I}_\nu$ from the prior in Fig.~\ref{fig:prior_wsmooth}(c) are more oscillatory than those in (f), for which the power spectra are smooth. However because this increased correlation corresponds to baselines at which the visibility amplitude is also small, the impact on the posterior mean brightness profiles in Fig.~\ref{fig:prior_wsmooth}(c) is small. This need not be the case generally, but is typical of the power spectra generated by \fr because these regions of low power have been determined from the visibilities. The power spectrum estimate  $\tilde{\vec{I}}_{\nu,\ {\rm C}43-6}^2$ in Fig.~\ref{fig:prior_wsmooth}(a) and (d) further demonstrates this relative insensitivity. The comparatively large discrepancy between it and $\vec{p}_{\rm MAP}$ in Fig.~\ref{fig:prior_wsmooth}(d) relative to that in (a) is a result of the power spectrum estimate using $w_{\rm smooth} = 10^{-1}$ having higher amplitudes than that using $w_{\rm smooth} = 10^{-4}$ and thus providing a weaker constraint on the reconstructed brightness profile, yet this does not correspond to a less accurate brightness profile reconstruction in (f) relative to that in (c).

\begin{table*}
\caption{Model hyperparameters, their default values, and reasonable ranges to optionally vary these (discussed in Sec.~\ref{sec:param_sensitivity}).}
\begin{tabular}{l c c}
    \hline
    Hyperparameter & Default value (suggested range) & Referenced in \\
    \hline
    {\bf Brightness profile hyperparameters}\\
    $N$, number of collocation points & 300 ($100 - 300$) & Text following Equation~\ref{eq:beta}\\ 
    $R_{\rm out}$, fit's maximum radius & $2\arcsec$ ($\gtrsim$1.5$\times$ disc's outer edge) & Equation~\ref{eq:fourier-bessel}\\
    \hline
    {\bf Maximum a posteriori power spectrum hyperparameters}\\
    $\alpha$, order parameter for the power spectrum's inverse $\Gamma$ distribution hyperprior & 1.05 (1.00 -- 1.30) & Equation~\ref{eq:psmooth}\\
    $p_0$, scale parameter for the power spectrum's inverse $\Gamma$ distribution hyperprior  & $10^{-15}$ Jy$^2$ ($> 0, \ll 1$) & Equation~\ref{eq:prior_p}\\
    $w_{\rm smooth}$, strength of smoothing applied to the power spectrum & $10^{-4}$ ($10^{-4} - 10^{-1}$) & Equation~\ref{eq:p_smooth_1}\\
    \hline
\end{tabular}
\label{tab:params} 
\end{table*}

\subsection{The reconstructed brightness profile's sensitivity to the hyperparameters}
\label{sec:param_sensitivity}
Our model for the deprojected brightness profile has five free hyperparameters: two specifying the radial points at which the profile is reconstructed, $N$ and $R_{\rm out}$; and three on the power spectrum, denoted collectively as $\boldsymbol{\beta} = \{\alpha, p_0, w_{\rm smooth}\}$, where $w_{\rm smooth} = 1 / \sigma_{\rm s}^2$ (analogous to the definition of the visibility weights). Definitions, default values and reasonable bounds to optionally vary these hyperparameters are summarized in {\bf Table~\ref{tab:params}}.
Here we motivate the default values and discuss the fit's sensitivity to these choices.

We have found the reconstructed brightness profile to be insensitive to $R_{\rm out}$ and $N$ so long as they are sufficiently large, and also to $p_0$ so long as it is small relative to the power in the visibilities ($\ll 1$). The default $R_{\rm out} = 2\arcsec$ is sufficiently large for a protoplanetary disc. Alternatively if the disc radius is known, $R_{\rm out}$ can be set to a value somewhat (e.g., 50\%) larger than this, the main constraint being that the model assumes the flux is 0 beyond $R_{\rm out}$. By default $N$ is set to 300 points to ensure that $Q_{\rm max}$ exceeds the maximum baseline in a dataset (which is important for fit stability).

\begin{table*}
\caption{Observational quantities for all \emph{mock} datasets. All mock observations observe at 230 GHz (1.3 mm) with 7.5 GHz continuum bandwidth.}
\begin{tabular}{l c c c c c}
    \hline
    Mock observational setup & ALMA  & Synthesized beam  & Integration  & Peak $I$ [$10^{10}$ Jy sr$^{-1}$] &  Used in\\
    & configuration & FWHM [mas] & time [min] & (integrated $I$  & \\
    & (all in Band 6) & (longest baseline [$10^6\ \lambda$]) & (sampling time [s]) & [$10^{10}$ Jy sr$^{-1}$]) & \\
    \hline
    {\bf Mock profiles shown in Fig.~\ref{fig:morphologies}}\\
    \ \ \ \ \ low resolution
    & C43-3 & $590 \times 700$ ($\approx$0.4) & 2 (10) & 1.00 (0.25 -- 0.51) & Fig.~\ref{fig:prior_wsmooth},~\ref{fig:uncertainty},~\ref{fig:morphologies},~\ref{fig:non_negative} \\
    \ \ \ \ \ moderate resolution
    & C43-6 & $130 \times 170$ ($\approx$2) & 10 (10) & 1.00 (0.25 -- 0.51) & Fig.~\ref{fig:regularization},~\ref{fig:prior_wsmooth},~\ref{fig:morphologies} \\
    \ \ \ \ \ high resolution
    & C43-6 + C43-9 & $24 \times 30$ ($\approx$10) & 10 + 50 (10) & 1.00 (0.25 -- 0.51) & Fig.~\ref{fig:prior_alpha} \\
    \hline
    {\bf Mock profile based on AS~209} \\
    \ \ \ \ \ low noise  
    & C43-7 & $86 \times 106$ ($\approx$3) & 40 (10) & 5.00 (0.87) & Fig.~\ref{fig:method},~\ref{fig:sensitivity}\\ 
    \ \ \ \ \ moderate noise  
    & C43-7 & $86 \times 106$ ($\approx$3) & 2 (10) & 5.00 (0.87) & Fig.~\ref{fig:sensitivity}\\ 
    \ \ \ \ \ high noise  
    & C43-7 & $86 \times 106$ ($\approx$3) & 1 (10) & 5.00 (0.87) & Fig.~\ref{fig:sensitivity}\\ 
    \hline
\end{tabular}
\label{tab:obs} 
\end{table*}

The hyperparameters $\alpha$ and $w_{\rm smooth}$ do not appear directly in Equation~\ref{eq:mean_cond}; their effects on the brightness profile reconstruction enter entirely through their effects on $\vec{p}_{\rm MAP}$. 

As motivated in Sec.~\ref{sec:pmap_as_prior}, increasing $w_{\rm smooth}$ reduces structure in the fit's power spectrum, reducing correlation in the brightness profile at scales for which the power spectrum has low local amplitude, while increasing correlation between adjacent radial collocation points. However we have demonstrated in Fig.~\ref{fig:prior_wsmooth} that varying $w_{\rm smooth}$ within sensible bounds often has a negligible effect on the \fr reconstructed brightness profile. 
As also discussed in Sec.~\ref{sec:pmap_as_prior}, increasing $\alpha$ beyond 1.0 increases the SNR threshold below which the power spectrum falls toward $p_0$, damping variations in the brightness profile on scales where the visibility SNR is low (primarily at unconstrained scales beyond a dataset's longest baseline). 

{\bf Fig.~\ref{fig:prior_alpha}} examines a fit's typical sensitivity to the value of $\alpha$. With $\alpha=1.00$ (not a recommended value, shown only for pedagogy), \fr will fit noise-dominated data, introducing noise into the reconstructed brightness, as in the Jeffreys prior fit in Fig.~\ref{fig:prior_alpha} that is placing significantly more power in baselines beyond $\approx$2 {\ml} than exists in the noiseless input visibilities. Increasing $\alpha$ to $1.05$ is a conservative choice, mildly damping the power on scales where the SNR is low (causing the corresponding fit in Fig.~\ref{fig:prior_alpha} to walk off the data at $\approx$1.5 {\ml}). This value of 1.05 is our default choice for $\alpha$ because many moderate to high resolution ($\gtrsim 0.1$ {\ml}) datasets for protoplanetary discs exhibit noise-dominated visibilities at their longest baselines (a consequence of sampling density decreasing strongly at the most extended ALMA configurations for typical integration times).
Further increasing $\alpha$ will more aggressively damp power on scales with low SNR, though in some cases this can lead to the fit attaining a lower effective resolution (if a significantly wider range of scales are damped). That said, comparing the $\alpha = 1.05,\ 1.10$ and $1.30$ fits in Fig.~\ref{fig:prior_alpha} shows that varying $\alpha$ within sensible bounds often has an insignificant effect on the brightness profile, especially if the visibilities' SNR is dropping rapidly at long baselines. For a scenario in which varying $\alpha$ has a more significant impact, we will consider a disc with sub-beam features sampled at low SNR over a wide range of baselines in Fig.~\ref{fig:sensitivity} (presented in Sec.~\ref{sec:snr}).

In our tests the effect on the brightness profile of varying $\alpha$ and/or $w_{\rm smooth}$ is thus often trivial. Nonetheless, as a precaution \emph{we recommend varying the hyperparameters used in a fit within sensible bounds} (see Table~\ref{tab:params}) \emph{to assess the brightness profile's resulting sensitivity.} 
We have found that especially for lower resolution datasets, setting $\alpha$ and/or $w_{\rm smooth}$ too high can average
over real, underresolved features, causing them to appear broader and shallower. Hence our default values for these hyperparameters are at the lower bound of our suggested ranges in Table~\ref{tab:params}.

\subsection{Model limitations}
\label{sec:model_limitations}
\begin{enumerate}
\item Our axisymmetric model is 1D, fitting for the azimuthal average of the visibility data at each spatial scale. In the presence of deviations from axisymmetry the model is thus biased. For mild asymmetries the effect is not severe, averaging over brightness asymmetries azimuthally. However for major asymmetries (such as a prominent spiral) and/or when $(u,v)$ coverage results in broad gaps over a given baseline range, the model can break down. This also holds for observations in which there is more than one source. 

\item Since a centered axisymmetric model has only real visibilities, we do not fit Im(V). 

\item The model makes the flat sky approximation (Equation~\ref{eqn:vis_fundamental}), which assumes that the observation's region of the sky is sufficiently small.

\item Before fitting for the brightness profile, the visibilities must first be deprojected and phase-centered. \fr can optionally do this by fitting a 2D Gaussian to the visibilities, though this deprojection operation may yield an erroneous result if a disc has an appreciable vertical thickness or if limb darkening from the optically thick surface is important.

\item The fitted brightness profile does not include primary beam correction. For sources that were observed close to the center of the primary beam, the correction is typically small and can be obtained by dividing the reconstructed brightness profile by the primary beam profile $\mathcal{A}(\theta)$, where $\theta$ is the radial angular coordinate.

\item Regions in the visibilities with sparse and/or sufficiently noisy sampling can cause a lack of constraint on the local spatial frequency scale, inducing oscillations in the brightness profile on the corresponding spatial scale. This can potentially mimic real structure. The model typically prevents this by damping power on scales with low SNR, but when it does occur the oscillations in the brightness profile can be diagnosed by their frequency, which corresponds to the unconstrained spatial frequency scale. \emph{Varying the hyperparameter values for a fit as noted in Sec.~\ref{sec:param_sensitivity} is useful to assess and potentially suppress this behavior.} 

\item The uncertainty on the fitted brightness profile is typically underestimated. For this reason we do not show the uncertainty on \fr brightness profiles in this work. The model framework produces an estimate of the uncertainty on the brightness profile ($\diag{\mat{D}}$), but this is not reliable because reconstructing the brightness from Fourier data is an ill-posed problem. For example if the visibility amplitude were to spike at any point beyond the data's maximum baseline, this would imprint high amplitude variations in the brightness profile on small spatial scales. Unless we know a priori (which is not generally the case) that the visibilities are decreasing sufficiently rapidly with increasing baseline, the uncertainty is therefore formally infinite. While it is reasonable to assume that for real disc brightness profiles the visibilities do decrease rapidly at long baseline, it is not straightforward to generically extrapolate the slope of this decline beyond a dataset's longest baseline; a robust error estimate is thus difficult to obtain.

\begin{figure}
	\includegraphics[width=\columnwidth]{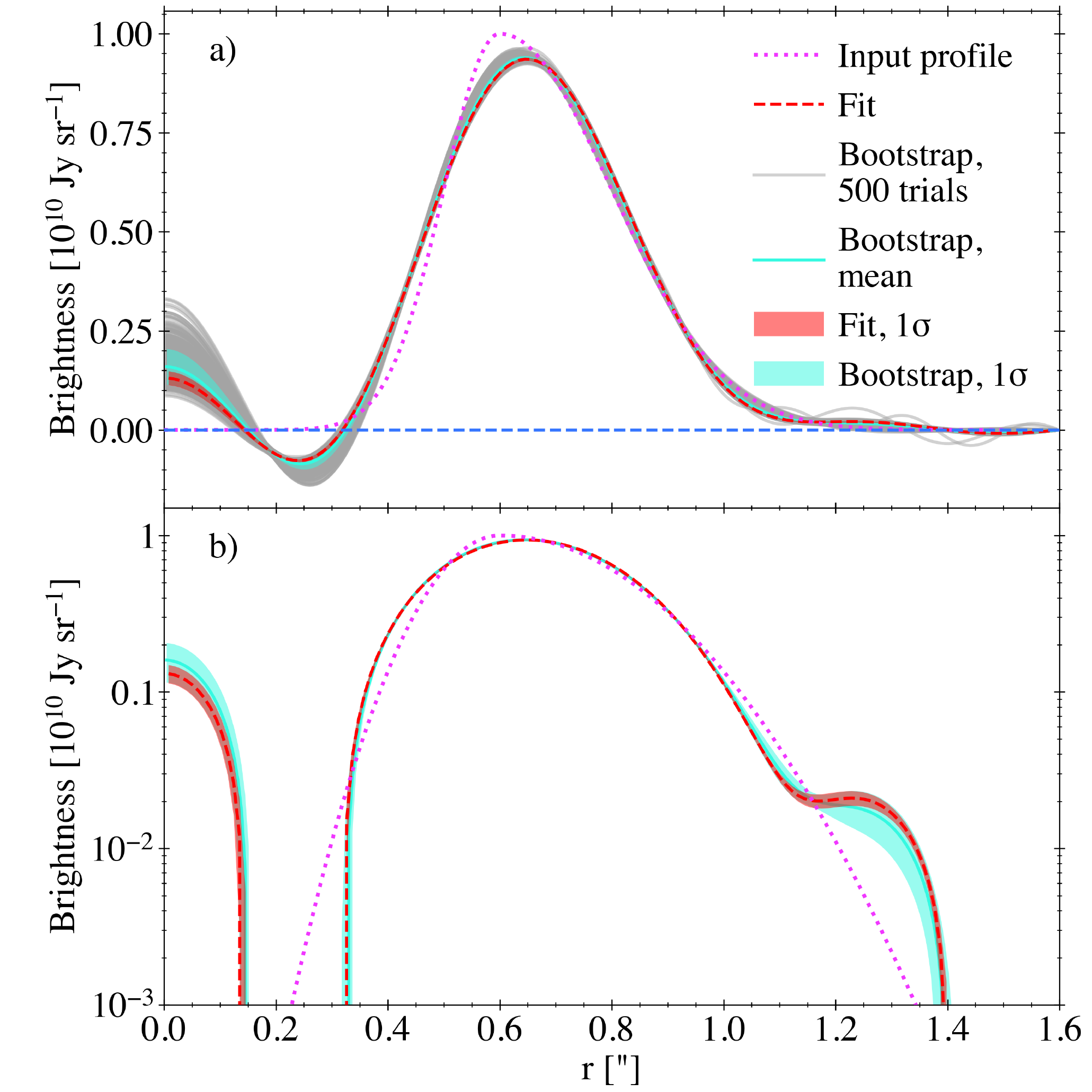}
	    \caption{{\bf Underestimated model uncertainty} 
	    \newline
	    a) Input and reconstructed brightness profiles for a mock Gaussian ring (two joined sigmoids) observed with the ALMA C43-3 configuration (synthesized beam FWHM $0.59 \times 0.70 \arcsec$, Briggs=0.5; see Table~\ref{tab:obs}). The fit's $1\sigma$ uncertainty estimate, $\diag{\mat{D}}$ estimated at the maximum a posteriori power spectrum, is shown. Additionally shown are 500 realizations of bootstrapping on this mock dataset, as well as the resulting distribution's mean and $1 \sigma$ uncertainty. Both the \fr fit's uncertainty and the bootstrap uncertainty are clearly underestimated as discussed in Sec.~\ref{sec:model_limitations}. \newline
	    b) As in (a) on a logarithmic scale.
        }
    \label{fig:uncertainty}
\end{figure}

{\bf Fig.~\ref{fig:uncertainty}} shows the \fr uncertainty estimate, $\diag{\mat{D}}$ estimated at the maximum a posteriori power spectrum, for the mock Gaussian ring presented in Sec.~\ref{sec:intrinsic_capability}. This confidence interval approximately represents the fit's statistical uncertainty (that due to the uncertainty on the observed baselines), which is correct if the visibility weights are an accurate representation of the pointwise visibility uncertainty. But the confidence interval does not capture the fit's ill-defined systematic uncertainty (that due to sparse sampling in the $(u,v)$ plane). A bootstrap on the visibilities in Fig.~\ref{fig:uncertainty} also fails to yield a reasonable estimate of the systematic uncertainty. Though this does confirm that $\diag{\mat{D}}$ is a reasonable estimate of the statistical uncertainty. \emph{We therefore urge caution before using uncertainty estimates to interpret the significance of the result in the recovered profile.}

To test whether including the uncertainty on $\vec{p}$ has a significant effect on the final uncertainty of the reconstructed brightness, we used the estimate for the uncertainty on \vec{p} given by \citet{2013PhRvE..87c2136O}. We translated the effect of this on the brightness profile by Taylor expanding $P(\vec{V}, \vec{p})$ about its maximum to make a Gaussian approximation to $P(\vec{V}, \vec{p})$. Using this approximation, we then drew samples and compared the variance of the reconstructed brightness at each radial collocation point to that estimated by $\diag{\mat{D}}$. Except for an uninformative Jeffreys prior ($\alpha=1.0$ and $w_{\rm smooth}=0$), the effect was negligible. 

\item The fitted brightness profile can have negative regions corresponding to spatial scales un- or underconstrained by the visibilities. There is an argument for choosing a fitting strategy that enforces the solution be nonnegative (as in \citealt{2016A&A...586A..76J}), and we have investigated the effect of negative fit regions by finding the most probable nonnegative intensity profile given $\vec{p} = \vec{p}_{\rm MAP}$. The effect on the recovered brightness profile is localized to the regions of negative flux, with otherwise minor differences. We explore a pedagogical nonnegative fit in Appendix~\ref{sec:non_negative}. 
\end{enumerate}

\subsection{Code performance} 
\label{sec:performing_fit}
The code's computation time is dominated by two components, constructing the matrix $\mat{M}$ and iterating the fit. The construction of $\mat{M}$ is only done once at the start of the fit and has computational cost  $\mathcal{O}(N_{\rm collocation\ points}^2 N_{\rm visibilities})$, while solving the linear systems in each iteration scales as $\mathcal{O}(N_{\rm collocation\ points}^3) \sim 100^3$.

To limit memory requirement, the matrix $\mat{M}$ is assembled in blocks, avoiding the need to hold the $N_{\rm collocation\ points} \times  N_{\rm visibilities}$ matrix $\mat{H}(\vec{q})$ in memory. Typical computational requirements were estimated using real datasets: with $10^4$ visibilities and 200 collocation points the fit took 10 s and used $\approx$100 MB, while with $10^6$ visibilities it took 40 s and used $\approx$200 MB. These tests were conducted on a 2017 MacBook Pro with a 7th generation Intel Core i5 processor (7360U) running at 2.3 GHz with 8 GB RAM.

\section{Demonstration \& analysis}
\label{sec:results}
\subsection{Demonstration on mock observations}
\label{sec:intrinsic_capability}
\subsubsection{Fits can attain sub-beam resolution}
\label{sec:convolution}
\begin{figure*}
	\includegraphics[width=\textwidth]{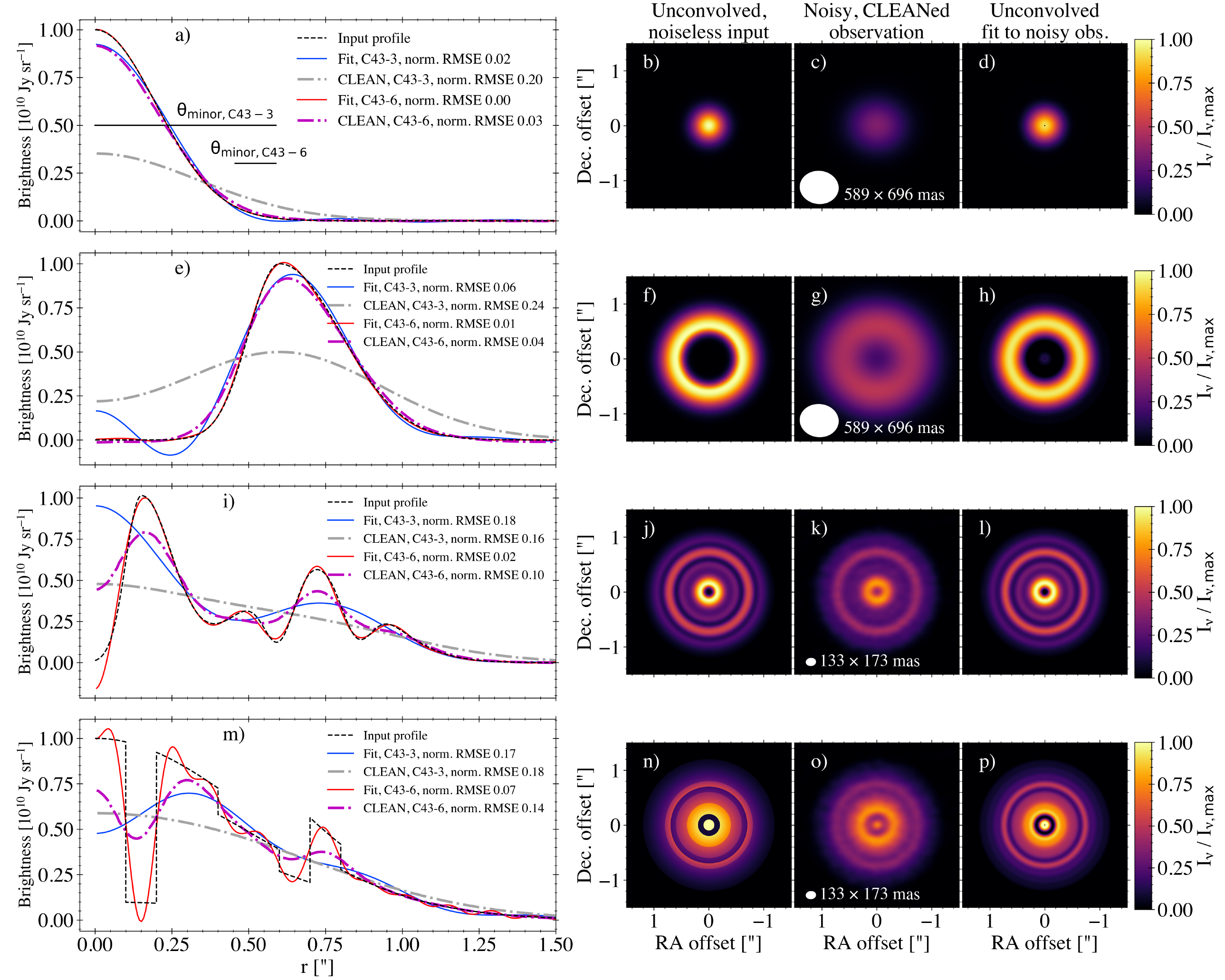}
	    \caption{{\bf Performance on different disc morphologies using mock observations} 
	    \newline
	    a) Input and reconstructed brightness profiles for a mock Gaussian disc with FWHM $0.2 \arcsec$, observed separately with the ALMA C43-3 (synthesized beam FWHM $0.59 \times 0.70 \arcsec$, Briggs=0.5) and C43-6 (beam FWHM $0.13 \times 0.17 \arcsec$, Briggs=0.5, factor of 5 longer integration time than C43-3; see Table~\ref{tab:obs}) configurations. The beams' minor axes are shown for reference. The \fr fit to each of the 2 mock observations is shown, as are CLEAN image-extracted profiles. The normalized RMS error of each profile is given. \newline
	    b) Input profile swept over $2 \pi$, noiseless and at infinite resolution. \newline
	    c) Mock CLEAN image (C43-3 + noise), with brightness normalized to (b). \newline
	    d) Image of the \fr fit to the noisy C43-3 mock observation, $not$ convolved with the beam, with brightness normalized to (b). \newline
        e) -- h) As in (a) -- (d) but for a Gaussian ring (two joined sigmoids) with width $\approx 0.4 \arcsec$. \newline
	    i) -- l) As in (a) -- (d) but for a more complicated, smooth-featured disc. The feature widths vary from $0.10 - 0.15 \arcsec$. The images correspond to the C43-6 observation. \newline
        m) -- p) As in (i) -- (l) but for a disc with step features, the narrowest of which is $0.1 \arcsec$.
        }	    
    \label{fig:morphologies}
\end{figure*}

To demonstrate \fr's fitting approach and characterize its performance, we begin in {\bf Fig.~\ref{fig:morphologies}} with a series of mock observations for discs of archetypal smooth and sharp structure at both well resolved and underresolved scales. We generate these mock datasets using the default fit hyperparameters in Table~\ref{tab:params} and the observational setups in {\bf Table~\ref{tab:obs}}.

Fig.~\ref{fig:morphologies}(a) compares the brightness reconstructed by \fr to the CLEAN image-extracted profile for a Gaussian disc centered at 0. The Gaussian's width is resolved by the CLEAN beam for the C43-6 ALMA configuration, yet importantly the CLEAN profile is slightly too broad and shallow as a result of beam convolution\footnote{As a simple illustration of this effect, noiseless Gaussians with FWHM 25, 50, 75, and 100 mas would respectively be broadened by convolution with a FWHM $50 \times 50$ mas Gaussian beam to FWHM 56, 71, 90, and 112 mas (a factor of 2.24, 1.42, 1.20, and 1.12 increase), and their amplitudes would be reduced to 45, 70, 83 and 89\% their true values.}. By comparison \fr recovers the profile to $<$1\% RMS error (note this is the standard definition of RMS error, unrelated to the RMS noise in a CLEAN image). For the same disc observed at the lower resolution C43-3 configuration, the beam's minor axis underresolves the Gaussian by a factor of $2.3$, and the effect of beam convolution on the CLEAN profile is exacerbated. The \fr fit by comparison retains high accuracy, comparable to the CLEAN profile from the C43-6 observation. 
Fig.~\ref{fig:morphologies}(b) -- (d) show the 2D images for the C43-3 case, with the \fr image recovering the input image from the noisy, low resolution observation.
This simple Gaussian case illustrates that because the model does not at any stage require beam convolution, \texttt{frank}\emph{'s achievable resolution is sub-beam}.

Fig.~\ref{fig:morphologies}(e) -- (h) next consider a slightly more complex disc, a Gaussian ring. Here again the \fr fit to the C43-3 observation achieves a similar accuracy to the CLEAN profile extracted from the higher resolution C43-6 dataset. 
This is despite the C43-3 beam's minor axis underresolving the Gaussian by a factor of $\approx$1.5. The \fr profile reconstructed from the C43-3 dataset does however misidentify the Gaussian's centroid and also shows a region of negative brightness allowed by the model as detailed in Sec.~\ref{sec:model_limitations}. We discuss a fit with enforced brightness positivity for this disc in Appendix~\ref{sec:non_negative}. 

As well as reproducing simple profiles, \fr adapts effectively to more complicated discs such as that in Fig.~\ref{fig:morphologies}(i) -- (l). Here the C43-3 beam's minor axis is a factor of $>$3 broader than the profile's widest feature, and \fr is unable to reconstruct the profile accurately. However it does discern that there are two well-separated peaks, while the CLEAN profile does not show any substructure. Increasing the resolution to C43-6, \fr retains its resolving power advantage, recovering the input brightness profile to high accuracy. 
 
To strain the model, Fig.~\ref{fig:morphologies}(m) -- (p) introduces perfectly sharp-edged features (step functions) to be recovered. \fr fits this disc to reasonable fidelity in the C43-6 case but does show some oscillations. These are a consequence of Gibbs phenomenon, which arises when representing an infinitely sharp feature in Fourier space. By comparison the CLEAN profiles at both resolutions do not show these oscillations, but do smear the disc features over the beam. \fr's ability to recover arbitrarily sharp features with comparatively low error demonstrates utility in more accurately recovering a disc's often steep outer edge and its peak flux.  

Together these mock discs show \fr's ability to fit smooth and sharp, partially and well-resolved, faint and bright features, at sub-beam resolution. In practice this can enable similar fit resolution to a CLEAN profile obtained with a more extended array configuration (e.g., the \fr fit to the C43-3 data in Fig.~\ref{fig:morphologies}(a) is comparable to the CLEAN profile for the C43-6 data). The \fr fits for all 8 mock datasets shown in Fig.~\ref{fig:morphologies} recover the discs' total flux to within a mean $0.8\%$ (standard deviation $0.2\%$), compared to a mean $1.6\%$ (standard deviation $1.6\%$) for the CLEAN profiles. This error in total flux recovery increases as a disc's features become increasingly sub-beam, an effect that is more severe for CLEAN than \fr. All \fr fits shown here are negligibly sensitive to the choice of hyperparameter values within the suggested ranges listed in Table~\ref{tab:params}. As discussed in Sec.~\ref{sec:performing_fit} these fits -- and all others shown in this work -- are performed in $\lesssim$1~min, and the computation speed is independent of the complexity of disc substructures; \fr \emph{fits simple and complicated disc profiles equally fast.} 
                   
\begin{figure*}
	\includegraphics[width=\textwidth]{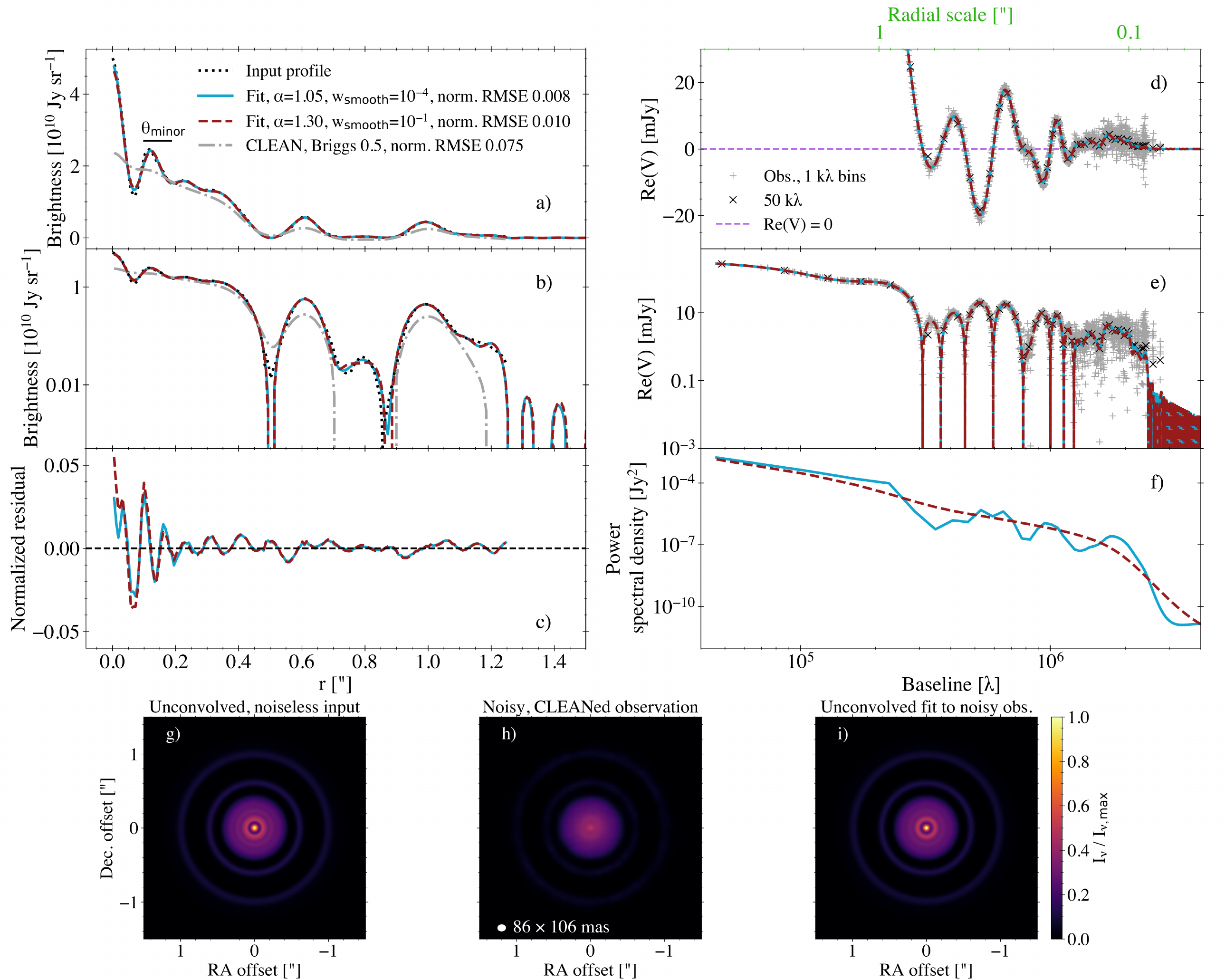}
	    \caption{{\bf Fit methodology using mock observations} 
	    \newline
	    a) Input and reconstructed brightness profiles under different hyperparameter values for a mock disc that uses as input the CLEAN image-extracted fit to the real DSHARP AS~209 observations (ALMA configurations C40-5/8/9, synthesized beam FWHM $36 \times 38$ mas; \citealt{Andrews2018}). The mock observation uses only the C43-7 configuration (beam FWHM $86 \times 106$ mas, Briggs=0.5; see Table~\ref{tab:obs}). This beam's minor axis is shown for reference. The CLEAN image-extracted profile for the mock C43-7 observation is also shown, with the normalized RMS error of all fits given. \newline
	    b) As in (a) on a logarithmic y-scale. The fit's oscillations beyond the disc's outer edge indicate the model's noise floor. \newline
	    c) Normalized residual for the \fr fits under different hyperparameter values. \newline
	    d) Visibilities for the mock C43-7 observation in (a). The brightness profile fits in (a) are the discrete Hankel transforms of these visibility domain fits. \newline
	    e) As in (d) on a logarithmic y-scale. \newline
	    f) The fit's maximum a posteriori power spectrum $\vec{p}_{\rm MAP}$ for different hyperparameter values. These power spectra are used as the priors on the brightness profile reconstructions in (a). \newline
	    g) Input profile swept over $2\pi$, noiseless and at infinite resolution. \newline
	    h) Mock CLEAN image (C43-7 + noise), with brightness normalized to (g). \newline
	    i) Image of the \fr fit to the noisy C43-7 mock observation, \emph{not} convolved with the beam, with brightness normalized to (g).
	    }
    \label{fig:method}
\end{figure*}

To illustrate  how \fr is attaining a sub-beam fit resolution, {\bf Fig.~\ref{fig:method}} characterizes its performance using a mock profile based on observational data rather than a simple functional form. We use the CLEAN image-extracted fit to the real DSHARP observations (C40-5/8/9 $+$ archival short and moderate baseline datasets) of AS~209 \citep{Andrews2018} as the input profile to be recovered from mock observations. 
This profile was obtained in that work with a beam of FWHM $36 \times 38\ {\rm mas}$; we generate the mock data at a factor of $\approx$2.6 worse resolution using the C43-7 configuration (beam FWHM $86 \times 106\ {\rm mas}$, see Table~\ref{tab:obs}). Much of the structure in the profile is thus sub-beam in the mock observation.
The mock visibilities' SNR as a function of baseline is similar to the real dataset.

\fr accurately recovers the input profile's sub-beam features to within $1\%$ RMS error for the noiseless case in Fig.~\ref{fig:method}(a) -- (b). It shows minor difference between the fit under the default hyperparameter values and under values at the other extrema of our suggested range (Table~\ref{tab:params}) as shown in Fig.~\ref{fig:method}(c), despite the difference in prior structure in Fig.~\ref{fig:method}(f). The brightness profile recovery's high fidelity is a result of the model fitting the visibility distribution in Fig.~\ref{fig:method}(d) -- (e) to high accuracy. As in previous cases the CLEAN profile extracted from the C34-7 mock image underresolves these features. The \fr fit recovers the mock disc's total flux to within $0.6\%$, while the analogous CLEAN error is $30.4\%$. The nontrivial CLEAN error is primarily a result of the mock observations containing no baselines shorter than those in the ALMA C43-7 configuration; i.e., the nominal maximum recoverable scale for C43-7, Band 6 is 1.12$\arcsec$ \citep{alma_handbook_2019}, while the disc extends to 1.25$\arcsec$. By comparison, \fr is able to extrapolate the fit accurately to short baselines despite this lack of data (though this behavior may not always be robust).
In Fig.~\ref{fig:method}(f) the maximum a posteriori power spectrum $\vec{p}_{\rm MAP}$ has low power at the unconstrained spatial frequencies beyond the data's longest baseline, preventing large amplitude oscillations in the reconstructed brightness profile at equivalent spatial scales (these oscillations, damped, are seen in the residuals of Fig.~\ref{fig:method}(c)). 
In Fig.~\ref{fig:method}(g) -- (i) the 2D images of the input model, mock CLEANed observation and fit demonstrate the amount of structure \fr is recovering that is smeared out by the beam.

\emph{Beam convolution is the primary difference in achievable resolution between} \fr \emph{and CLEAN}. Convolving either \fr fit in Fig.~\ref{fig:method}(a) -- (b) with the 2D CLEAN beam yields a profile that is similar to the CLEAN profile. 
The CLEAN beam size does depend on the visibility weighting, with the choice affecting the resulting CLEAN profile. For the disc in Fig.~\ref{fig:method}, setting the Briggs robust parameter to $-2.0$ (approximating uniform weighting), $0.5$ and $2.0$ (natural weighting) sets the beam FWHM as $0.08\arcsec \times 0.10 \arcsec,\ 0.09\arcsec \times 0.10\arcsec$ and $0.10\arcsec \times 0.12\arcsec$. However the resulting profiles vary only slightly, with the RMSE changing at most by $3\%$, and each profile still underresolves the disc features relative to \fr. This highlights that \emph{\fr can recover disc features underresolved even by uniform weighting, while retaining high sensitivity.}

\subsubsection{Baseline-dependent signal-to-noise determines the achievable fit resolution}
\label{sec:snr}
\begin{figure*}
	\includegraphics[width=\textwidth]{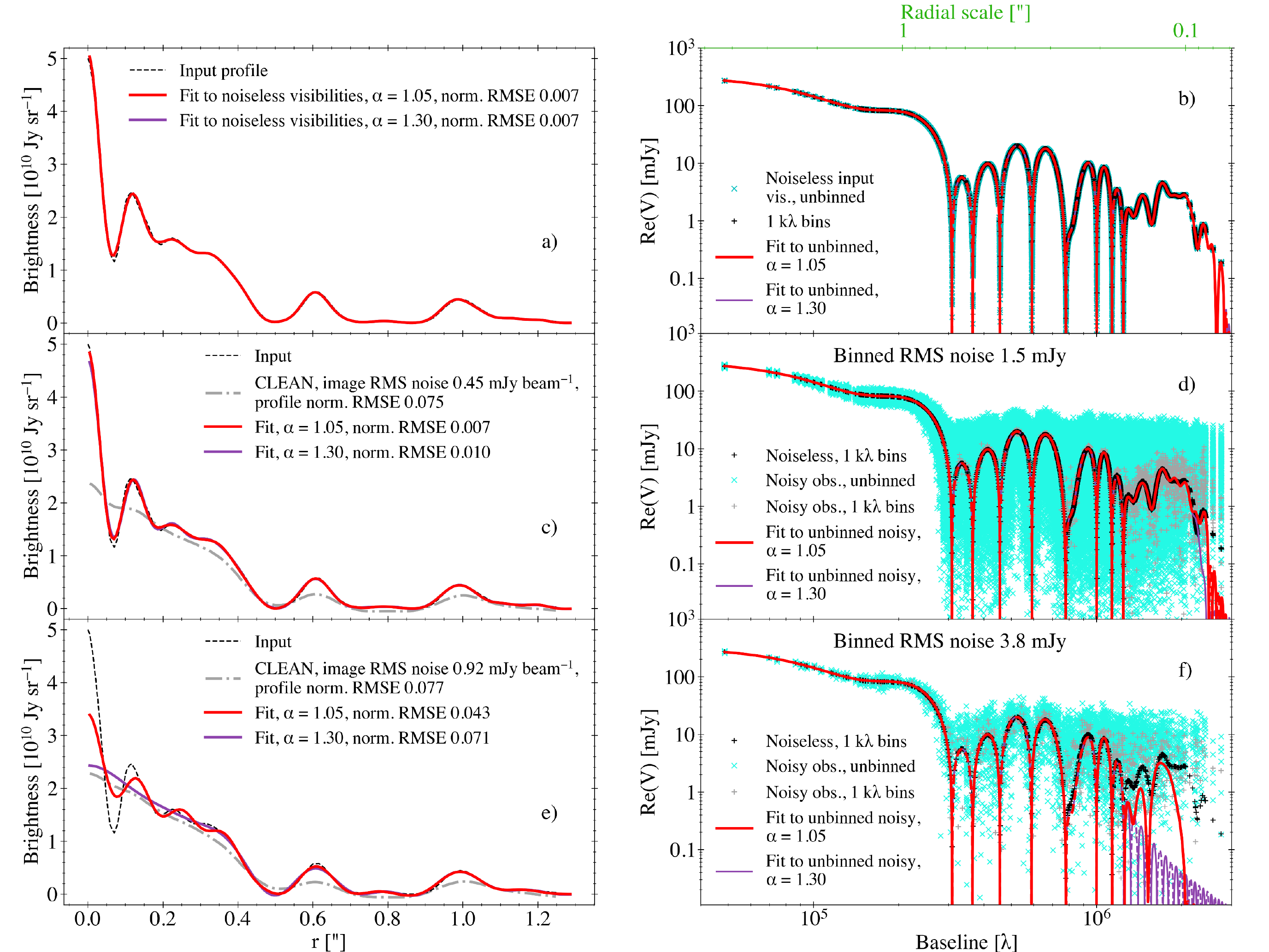}
	    \caption{{\bf Fit sensitivity to baseline-dependent SNR using mock observations} 
	    \newline
	    a) The input profile from Fig.~\ref{fig:method} is used to generate noiseless mock observations, with \fr fits to these noiseless data shown and their normalized RMS errors given under two values of the $\alpha$ hyperparameter (both fits use $w_{\rm smooth}=10^{-4}$). \newline
	    b) Noiseless visibilities corresponding to the input profile in (a). \fr fits the \emph{unbinned} visibilities shown. Also shown are the data in $1\ {\rm k}\lambda$ bins. The data (and the fits') negative regions are denoted by the fit lines' dashed sections. The $\alpha=1.30$ fit is behind the $\alpha=1.05$ fit. \newline
	    c) -- d) As in (a) -- (b) but with noise added to the mock observation. The fits shown are to the unbinned, noisy data. The CLEAN image-extracted profile is shown for comparison. The apparently larger scatter in the observations in (d) (compared with (f)) is due to the larger number of data points rather than higher intrinsic error. \newline
	    e) -- f) As in (c) -- (d) but with a higher RMS noise due to shorter integration time. The binned RMS noise (1 k$\lambda$ bins) does not increase dramatically because the number of empty bins has also increased. 
        }
    \label{fig:sensitivity}
\end{figure*}

We next characterize the model's SNR sensitivity by decrementing the integration time for mock observations of the input profile from Fig.~\ref{fig:method}. This increasingly degrades the $(u,v)$ coverage, in turn worsening the data's effective SNR at a given baseline. {\bf Fig.~\ref{fig:sensitivity}}(a) -- (b) first show the method's intrinsic capability by fitting the \emph{noiseless} data, with the $(u,v)$ coverage determined by the mock observation's C43-7 configuration and integration time. The fits using both values for the $\alpha$ hyperparameter accurately match the full visibility distribution, the reconstructed brightness profiles showing $<$1\% RMS error. 

Then fitting the same baseline distribution but with the mock observation's noise included in Fig.~\ref{fig:sensitivity}(c) -- (d), both fits remain able to recover the noiseless visibilities to high accuracy. However as the data's SNR worsens beyond $\approx$1.8 {\ml} the fits do show some error, including tending toward 0 prior to the longest baselines. These errors manifest in the recovered brightness profiles in Fig.~\ref{fig:sensitivity}(c) as a slightly less accurate recovery of the peak brightness and of the innermost gap's depth. The fit using the stronger $\alpha = 1.30$ is slightly less accurate at the longest baselines because the mock observations do not have sufficiently high SNR there for the model to fit them with $\alpha = 1.30$. However the differences in the reconstructed profile are minor for this case.

As the integration time is further decremented in Fig.~\ref{fig:sensitivity}(e) -- (f), the fits' fidelity is degraded primarily beyond $\approx$1.2 {\ml} as the visibility distribution becomes more sparse. By consequence the highest resolution structural information -- the depths and centroids of the narrowest features in the input brightness profile, including the peak brightness -- are only partially recovered. The effect is more severe for the $\alpha = 1.30$ case because the data beyond 1.2 {\ml} have low effective SNR, and as $\alpha$ increases, the SNR threshold below which \fr does not fit the visibilities increases. Since these data clearly contain meaningful information, $\alpha=1.05$ is the more sensible choice.

In contrast to \fr's SNR sensitivity, the CLEAN image-extracted profiles in Fig.~\ref{fig:sensitivity}(c) and (e) vary marginally when the visibilities' SNR is decreased. This is because convolution with the C43-7 beam, rather than the data's SNR, is primarily limiting the CLEAN image resolution. 
While \fr's resolving power is sensitive to variations in the baseline dependence of the data's SNR, the CLEAN profile is largely determined by the pure baseline distribution.
This entails that \emph{improving an observation's SNR via the on-source integration time can significantly enhance the resolving capability of} \fr, \emph{while it may make little difference for CLEAN.}

\subsection{Demonstration on real observations}
\label{sec:real data}
Mock datasets are useful to test and characterize \fr's performance, though real data have more complex noise properties to which the model must also be well suited.

\subsubsection{Sub-beam fits: Characterizing fine structure in real, high resolution observations}
\label{sec:real_data_hires}
\begin{figure*}
	\includegraphics[width=\textwidth]{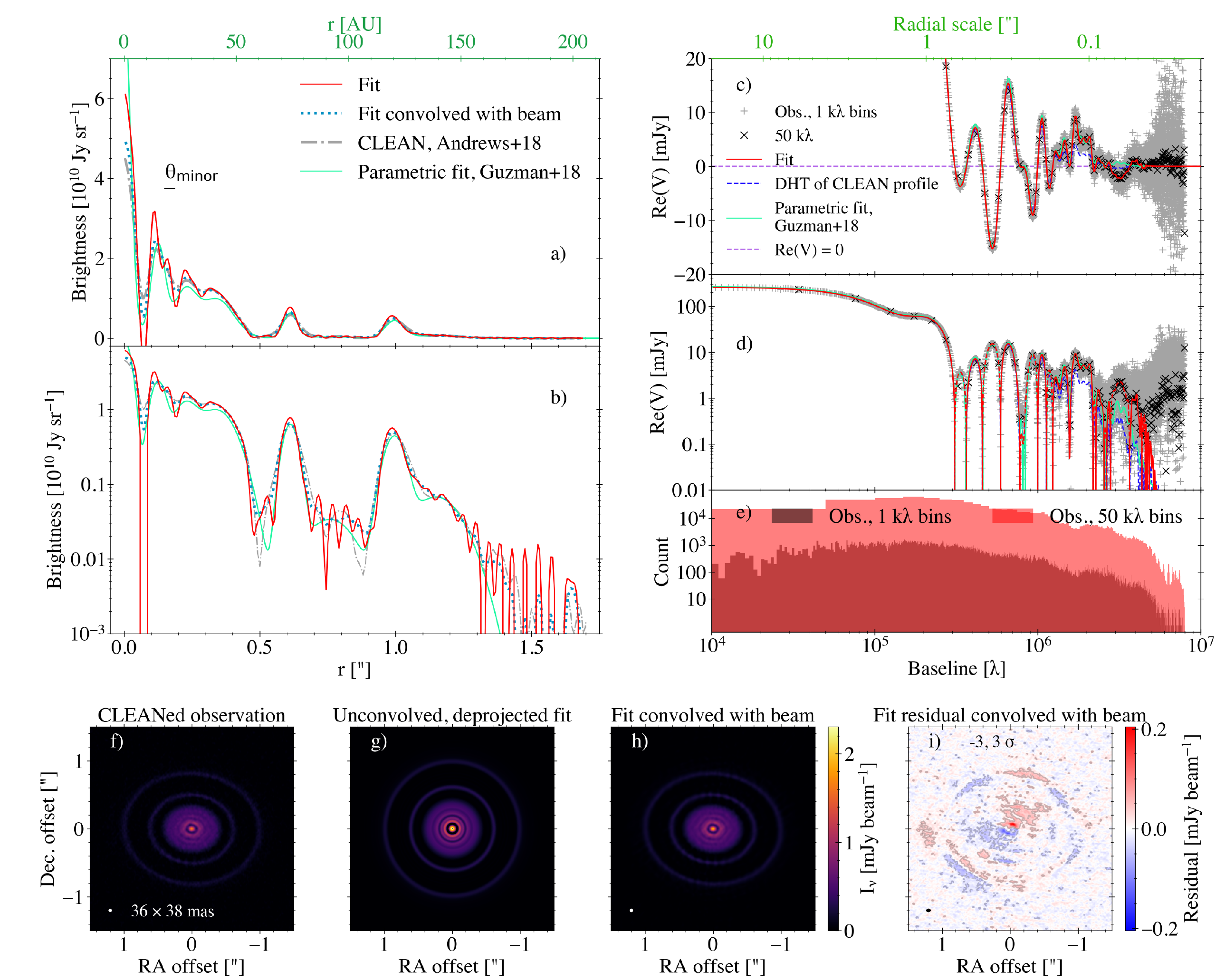}
	    \caption{{\bf Fit to real, high resolution observations: AS~209}
	    \newline
	    a) Reconstructed brightness profile for the real, high resolution (beam FWHM $36 \times 38$ mas) DSHARP observations of AS~209 in \citet{Andrews2018}. The CLEAN image-extracted profile from that work and the beam's minor axis are shown. Consistent with \fr demonstrations on mock data, the model achieves a higher fit resolution (narrower, higher amplitude features) than the CLEAN technique. The presence of additional structure in the inner disc relative to the CLEAN image-extracted profile is robust to variations of the model.
	    Also shown is the \fr brightness profile convolved with the observations' 2D synthesized beam and the brightness profile for the parametric visibility domain fit in \citet{Guzman2018}.
	    \newline
	    b) As in (a) on a logarithmic y-scale. The outermost radii show the fit's noise floor. \newline
	    c) Zoom on the region around 0 of the observed DSHARP visibilities in 1 and 50 {\kl} bins. The \fr fit, the discrete Hankel transform of the CLEAN profile in (a), and the parametric fit from \citet{Guzman2018} corresponding to the profile in (a) are shown. The strong increase in scatter at $\gtrsim 4.5$ {\ml} sets the data's \emph{effective} resolution, beyond which \fr does not attempt to fit the noise-dominated visibilities. \newline
	    d) As in (c) on a logarithmic y-scale. The data (and the \fr fit's) negative regions are denoted by the \fr fit line's dashed sections. \newline
	    e) Histogram of (binned) visibility counts, showing the strong decrease beyond $\approx 4.5$ {\ml} responsible for the increased scatter in (c) -- (d). \newline
        f) CLEANed observation. \newline
        g) Image of the \emph{unconvolved and deprojected} \fr reconstruction. \newline 
        h) Image of the \fr reconstruction reprojected and convolved with the observations' synthesized beam. \newline
        i) Residual map of the \fr reconstruction (the fit's residual visibilities imaged with CLEAN), showing evidence of non-axisymmetric structure (contour $\sigma = 10$ $\mu$Jy beam$^{-1}$, peak residual $12 \sigma$).
	    }
    \label{fig:as209_real}
\end{figure*}

To this end {\bf Fig.~\ref{fig:as209_real}} demonstrates \fr's performance with real data, fitting the high resolution DSHARP observations of AS~209 (synthesized beam FWHM $36 \times 38$~mas $\approx 5$ AU at 121~pc)\footnote{We downloaded the AS~209 self-calibrated and multi-configuration combined continuum measurement sets from 
\href{https://bulk.cv.nrao.edu/almadata/lp/DSHARP}{\color{linkcolor}https://bulk.cv.nrao.edu/almadata/lp/DSHARP}. Before extracting the visibilities using the \texttt{export\_uvtable} function of the \texttt{uvplot} package \citep{uvplot_mtazzari}, we applied channel averaging (to obtain 1 channel per spectral window) and time averaging (30 sec) to all spectral windows and multi-configuration datasets in the original MS table.}.

Fig.~\ref{fig:as209_real}(a) -- (b) show the \fr fit (using the default hyperparameter values) and the CLEAN image-extracted profile from \citet{Andrews2018}. \fr recovers additional substructure and higher amplitude features in the inner disc, a higher peak brightness, and slightly narrower rings in the outer disc. These results are consistent with those using mock observations in Sec.~\ref{sec:intrinsic_capability}. That the \fr fit finds a negative innermost gap indicates that the innermost disc's features are not fully resolved; forcing the fit to be positive results in the innermost gap having zero flux and alters the profile's integral by 4\%, but otherwise has a negligible effect (see Appendix~\ref{sec:non_negative} for a discussion of nonnegative fits).
The \fr fit agrees with the CLEAN profile in finding that the gaps centered at $\approx$0.5 and 0.8$\arcsec$ are not empty\footnote{Although the SNR in the gaps is low, this noise is dominated by the uncertainty on scales $\lesssim 0.05\arcsec$; it is the lower uncertainty on scales of the approximate gap widths, $0.1\arcsec$ and $0.3\arcsec$, that suggests the average flux in the gaps is nonzero.}. 
However the fine structure in these gaps remains uncertain because their brightness is near the fit's noise floor (which can be approximated by the amplitude of the oscillations beyond $\approx$1.3$\arcsec$). 

The \fr fit convolved with the synthesized beam is also shown in Fig.~\ref{fig:as209_real}(a) -- (b), its similarity to the CLEAN profile giving credence to \fr correctly finding real higher resolution structural information. The sub-beam resolution of the reconstructed brightness profile is enabled by \fr accurately fitting the visibilities in Fig.~\ref{fig:as209_real}(c) -- (d) out to $\approx$4.5 {\ml}. This is equivalent to an angular scale $\sim 1/(4.5$ {\ml}) = 46 mas = 5.5 AU, which is an indication of the data's effective resolution beyond which the visibilities are noise-dominated. In Fig.~\ref{fig:as209_real}(e) a histogram of the binned visibilities commensurately decreases sharply in counts beyond $\approx$4.5 {\ml}. Fitting longer baselines with our current modeling approach only imprints noise on the brightness profile as discussed in Sec.~\ref{sec:data_effective_res}. 

Although the CLEAN beam has a FWHM $36 \times 38$ mas that is less than the 46~mas equivalent of the \fr visibility fit, these two values are not directly comparable. A more direct comparison can be made in Fourier space; the DHT of the CLEAN profile in Fig.~\ref{fig:as209_real}(c) -- (d) demonstrates how convolution with the CLEAN beam effectively acts as a lowpass filter, downweighting the contribution to the CLEAN brightness profile from visibilities beyond $\approx$2.4 {\ml} $\Leftrightarrow$ 86 mas = 10.4 AU. This is the baseline at which the DHT of the CLEAN profile begins to show discrepancies with the observed visibilities. Convolution with the beam thus suppresses features in the profile on spatial scales smaller than 10.4 AU, causing them to appear broader and shallower than in the \fr reconstruction, which fits the visibilities out to $\approx$4.5 {\ml} $\Leftrightarrow$ 46 mas = 5.5 AU.  

The DHT of the CLEAN profile more generally shows a less accurate agreement with the visibilities than \fr beyond $\approx$1.5 {\ml}, motivating that the \fr brightness profile is correctly identifying high resolution structure.
For comparison the \emph{parametric} visibility domain fit in \citet{Guzman2018} uses the CLEAN image to motivate an 8 Gaussian functional form for the brightness profile, shown in Fig.~\ref{fig:as209_real}(a) -- (b). The Fourier transform of this parametric form in Fig.~\ref{fig:as209_real}(c) -- (d) accurately fits the visibilities out to $\approx$2.5 {\ml} but begins to show discrepancies at longer baselines. The brightness profile has less structure in the inner disc and lower amplitude features than the \fr fit. Because the \fr fit accurately traces the data out to $\approx$4.5 {\ml}, its higher achieved resolution finds the disc features to be narrower and higher in amplitude (though the \citealt{Guzman2018} fit is positive everywhere). This is an example of the comparative advantage of a nonparametric form to fit a complicated visibility distribution.

Fig.~\ref{fig:as209_real}(f) -- (i) compare the CLEAN image with the image of the \emph{unconvolved and deprojected} \fr fit, the fit convolved with the beam, and the convolved residual image of the fit. The latter demonstrates an additional use case of the model, identifying and isolating small deviations from axisymmetry (for large deviations from axisymmetry, our axisymmetric fits may not be reliable). Here the $5 - 10\%$ deviations in the brightness around each ring may potentially be explained by the gaps and rings being produced by the combination of a planet and a low viscosity as suggested in \citep{Fedele2018}. In such a case these asymmetries may be expected as a result of the low viscosity \citep{2020MNRAS.491.5759H}. 

\begin{figure}
	\includegraphics[width=\columnwidth]{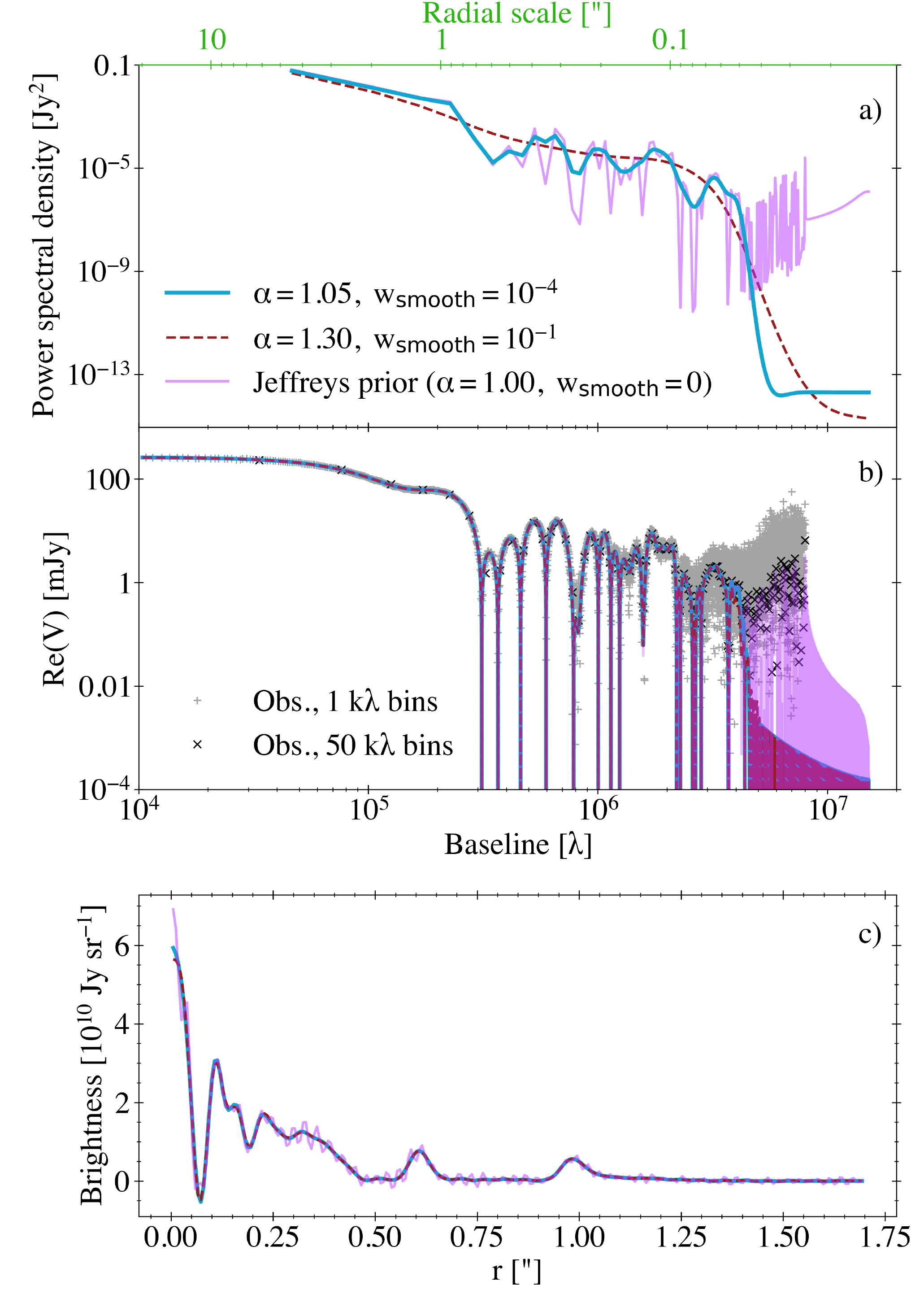}
	    \caption{{\bf Fit sensitivity to hyperparameters: AS~209}
	    \newline 
	    a) Maximum a posteriori power spectra of the \fr fit under the default hyperparameter values ($\alpha = 1.05,\ w_{\rm smooth} = 10^{-4}$) and stronger values ($\alpha = 1.30,\ w_{\rm smooth} = 10^{-1}$). The structural differences between these spectra show negligible effect on the visibility fits in (b) and fitted profiles in (c). The power spectra are the priors placed on the respective brightness profile reconstructions. The maximum a posteriori power spectrum under a Jeffreys prior ($\alpha = 1.00,\ w_{\rm smooth} = 0$) is also shown, which forces the model to fit the longest baseline, noise-dominated visibilities in (b), imprinting oscillations on the corresponding brightness profile in (c). \newline
	    b) Observed visibilities in 1 and 50 k$\lambda$ bins; the \fr fit under the default hyperparameter values and under a stronger choice, showing minor difference; the fit under a Jeffreys prior. \newline
	    c) The fitted \fr brightness profile in Fig.~\ref{fig:as209_real}(a) -- (b) under the default hyperparameter values, as well as fits under the stronger choice and Jeffreys prior.
	    }
    \label{fig:as209_prior}
\end{figure}

Varying the fit hyperparameters $\alpha$ and $w_{\rm smooth}$ within sensible bounds has negligible effect on the \fr brightness profile. {\bf Fig.~\ref{fig:as209_prior}} compares the fit from Fig.~\ref{fig:as209_real} using the default values ($\alpha = 1.05, w_{\rm smooth} = 10^{-4}$) with a fit that more strongly damps low SNR data ($\alpha = 1.30, w_{\rm smooth} = 10^{-1}$). The latter smooths the power spectrum appreciably relative to the default choice in Fig.~\ref{fig:as209_prior}(a), yet the effects on the visibility fit and the reconstructed brightness in Fig.~\ref{fig:as209_prior}(b) -- (c) are negligible; the fit is robust to how precisely the prior weights the longest (i.e., noisiest) baselines. Note that although the priors (power spectra) shown in Fig.~\ref{fig:as209_prior}(a) do not extend to the shortest baselines, this does not significantly impact the fit. This is because the maximum recoverable scale of the observations is much larger than the size of the disc, and \fr recovers the brightness accurately so long as $R_{\rm out}$ is larger than the disc size.

\begin{figure}
	\includegraphics[width=\columnwidth]{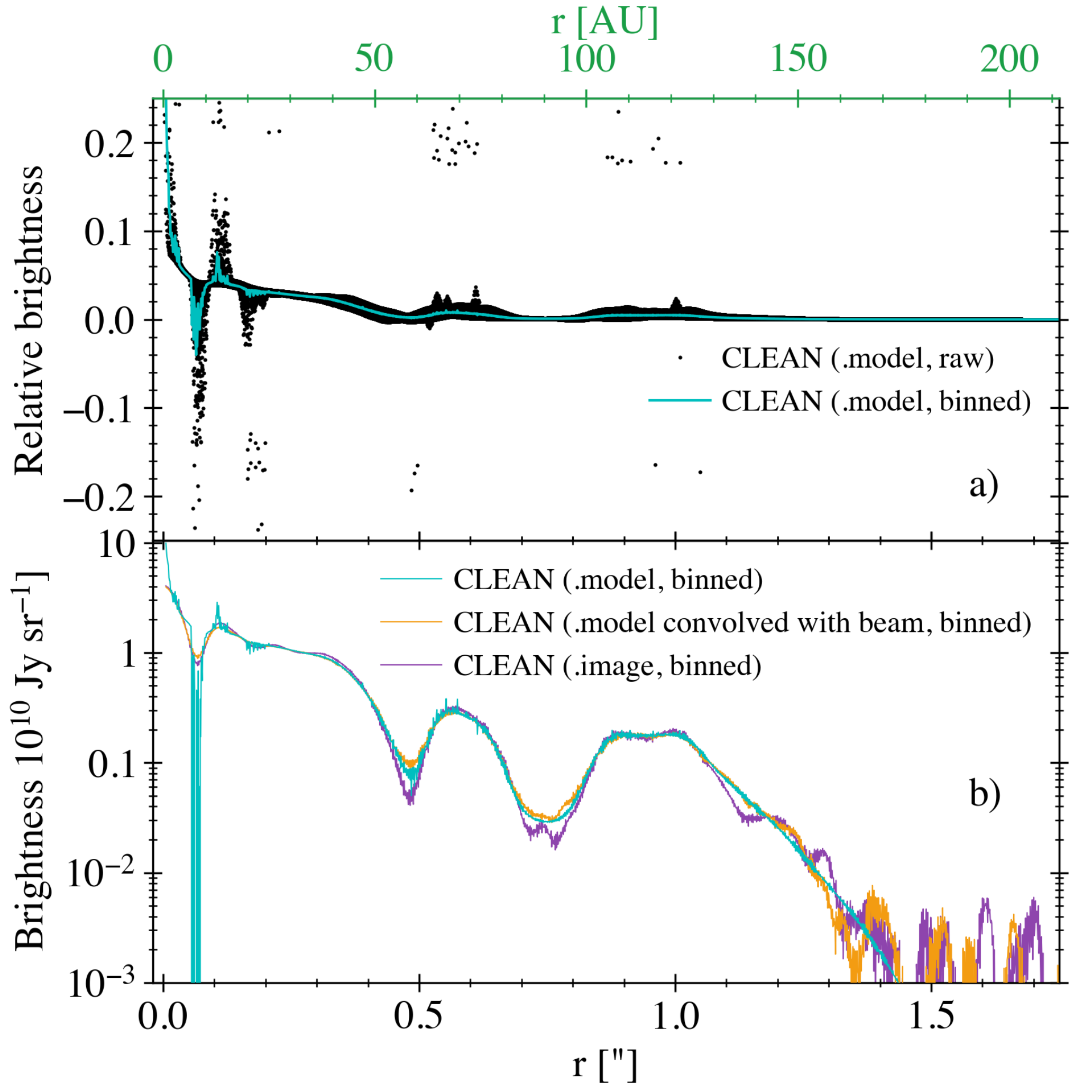}
	    \caption{{\bf CLEAN model profile extraction: AS~209} 
	    \newline
	    a) Pixel flux as a function of disc radius, extracted from the CLEAN point source model ({\it .model} image) for the DSHARP AS~209 dataset. A binned average (1 mas bins) of these data is also shown to estimate the radial profile underlying the noisy pixel flux. \newline
	    b) Binned average from (a), compared with a profile extracted (using the same binning) from the point source model convolved with the CLEAN beam ({\it .image - .residual}), and a profile extracted from the final CLEAN image ({\it .image}). The lack of prominent additional disc features in the {\it .model} profile indicates a comparable resolution to the {\it .image} profile.
	    }
    \label{fig:as209_clean}
\end{figure}

Although convolving the \fr profile with the CLEAN beam generally results in brightness profiles that are similar to CLEAN-extracted profiles, the converse is not true. In {\bf Fig.~\ref{fig:as209_clean}} we compare a profile extracted from the raw (unconvolved) CLEAN point source model ({\it .model} image); a profile extracted from the point source model convolved with the CLEAN beam (without the addition of the residual dirty map, the {\it .residual} image); and a profile extracted from the final CLEAN {\it .image} image.
One may expect that the fit resolution achieved in the CLEAN point source model profile is similar to that in the \fr profile. While the disc's innermost gap is narrower in the point source model profile relative to the final CLEAN image profile, the point source model profile does not show as much structure in the inner disc as the \fr fit. Moreover the noise in the point source model profile (even when binning) can make inference on disc feature widths and amplitudes nontrivial. We have applied this same analysis to multiple real and mock datasets at various antenna configurations, finding in general that while the resolution difference between profiles extracted from the {\it .model} image and {\it .image} image is starker in lower resolution data (because the beam is larger and so convolution with it has a stronger effect), the point source model profile's noise also worsens for lower resolution data. We have confirmed this behavior with a CLEAN gain parameter as low as $10^{-3}$.

\subsubsection{A dataset's effective resolution can be much less than the longest baseline}
\label{sec:data_effective_res}
To further demonstrate that the longest baselines in datasets are often noise-dominated, we find that both the \fr and CLEAN brightness profiles change negligibly if the AS~209 dataset, which extends to 8 {\ml} $\Leftrightarrow$ 26 mas = 3.1 AU, is truncated at 5 {\ml} $\Leftrightarrow$ 41 mas = 5.0 AU \emph{prior to} the visibilities being fit. This suggests that the data's effective resolution as seen by both CLEAN and \fr is $\lesssim5$ {\ml}, or at $\lesssim$63\% of the baseline distribution. In Fig.~\ref{fig:as209_real}(c) the visibilities begin oscillating rapidly about 0 beyond $\approx$5 {\ml}, indicative of noise dominating the signal. Fitting longer baselines only imposes noise on the reconstructed brightness profile as shown by the Jeffreys prior fit in Fig.~\ref{fig:as209_prior} (this behavior is consistent with that found using mock data in Fig.~\ref{fig:prior_alpha}). The Jeffreys prior causes the model to fit the full baseline distribution, including the noise-dominated region beyond $\approx 4.5$ {\ml}. This imposes oscillations on the brightness profile, demonstrating that these baselines in the AS~209 dataset are noise-dominated. The fit with the default hyperparameter values is thus (at least approximately) correctly identifying where the data become noise-dominated and accordingly justified in not fitting beyond $\approx$4.5 {\ml}. This is not to say that an alternative fitting approach could not obtain useful information on disc structures from these noise-dominated data.

Importantly though \emph{it is common for an appreciable fraction of the long baselines in real datasets to be noise-dominated}. This occurs in high as well as low -- moderate resolution observations and suggests the integration time at the most extended baselines is often too short to yield informative visibility measurements for CLEAN or \fr. This \lq{}effective resolution\rq{} is predominantly a result of sampling density dropping appreciably at the longest baselines in many real datasets (while the longest baselines also typically exhibit higher phase noise, self-calibration can often alleviate this).

\subsubsection{Sub-beam fits: Identifying underresolved features in real, moderate resolution observations}
\label{sec:real_data_midres}
\begin{figure}
	\includegraphics[width=\columnwidth]{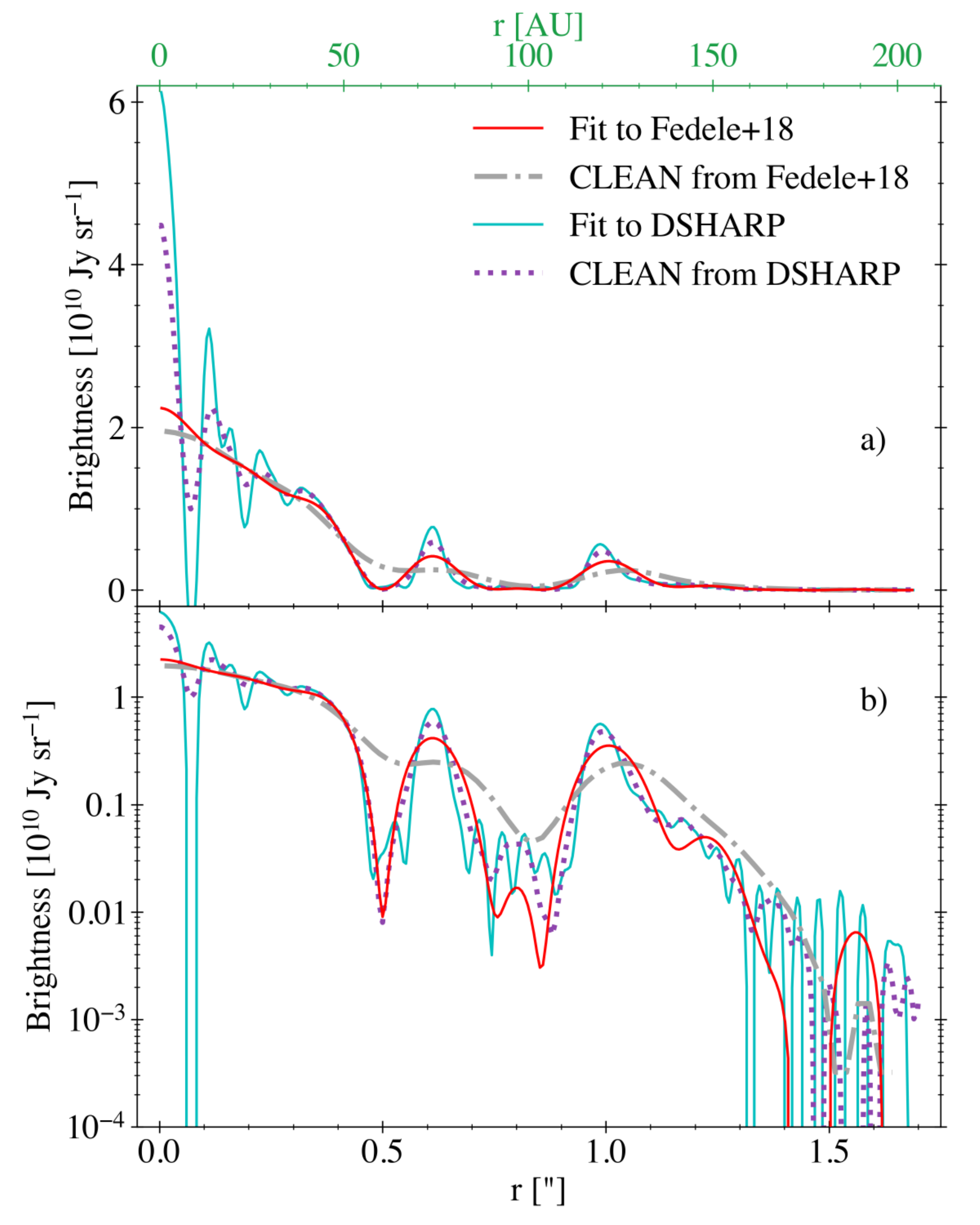}
	    \caption{{\bf Comparison of moderate and high resolution fits: AS~209}
	    \newline
	    a) The fitted \fr brightness profile in Fig.~\ref{fig:as209_real}(a), as well as a fit to lower resolution (longest baseline 2 {\ml}) observations of the same disc \citep{Fedele2018}. Analogs for the CLEAN image-extracted profile. The \fr fit to the lower resolution dataset is correctly identifying structure evident in both the \fr and CLEAN fits to the higher resolution data. \newline
	    b) As in (a) on a logarithmic y-scale.
	    }
    \label{fig:as209_lores}
\end{figure}

While \fr has utility in more accurately characterizing disc substructure in high resolution observations, it is also effective in identifying sub-beam structure in low -- moderate resolution observations. {\bf Fig.~\ref{fig:as209_lores}} compares the \fr and CLEAN fits to the DSHARP AS~209 dataset from Fig.~\ref{fig:as209_real} with fits to lower resolution (longest baseline 2 {\ml}) observations of the same disc \citep{Fedele2018}. Relative to CLEAN, the \fr fit to the lower resolution dataset is more accurately identifying a number of features in the brightness distribution (the centroids and widths of the outer disc's rings, depths of the adjacent gaps, and the disc's outer edge) as confirmed by the higher resolution DSHARP observations. The \fr fit to the lower resolution data also shows reasonable agreement with the CLEAN fit to the high resolution DSHARP data in the outer disc. This is due to both profiles fitting the visibilities out to a similar maximum baseline (the \fr fit to the lower resolution dataset extends to $\approx$2.1 {\ml}, and recall the DHT of the DSHARP CLEAN beam's FWHM is $\approx$2.4 {\ml}). \fr's sub-beam resolving capability is thus not limited to high resolution observations, offering the potential to identify and more accurately constrain substructure relative to CLEAN in low -- moderate resolution datasets.  

\begin{figure*}
	\includegraphics[width=\textwidth]{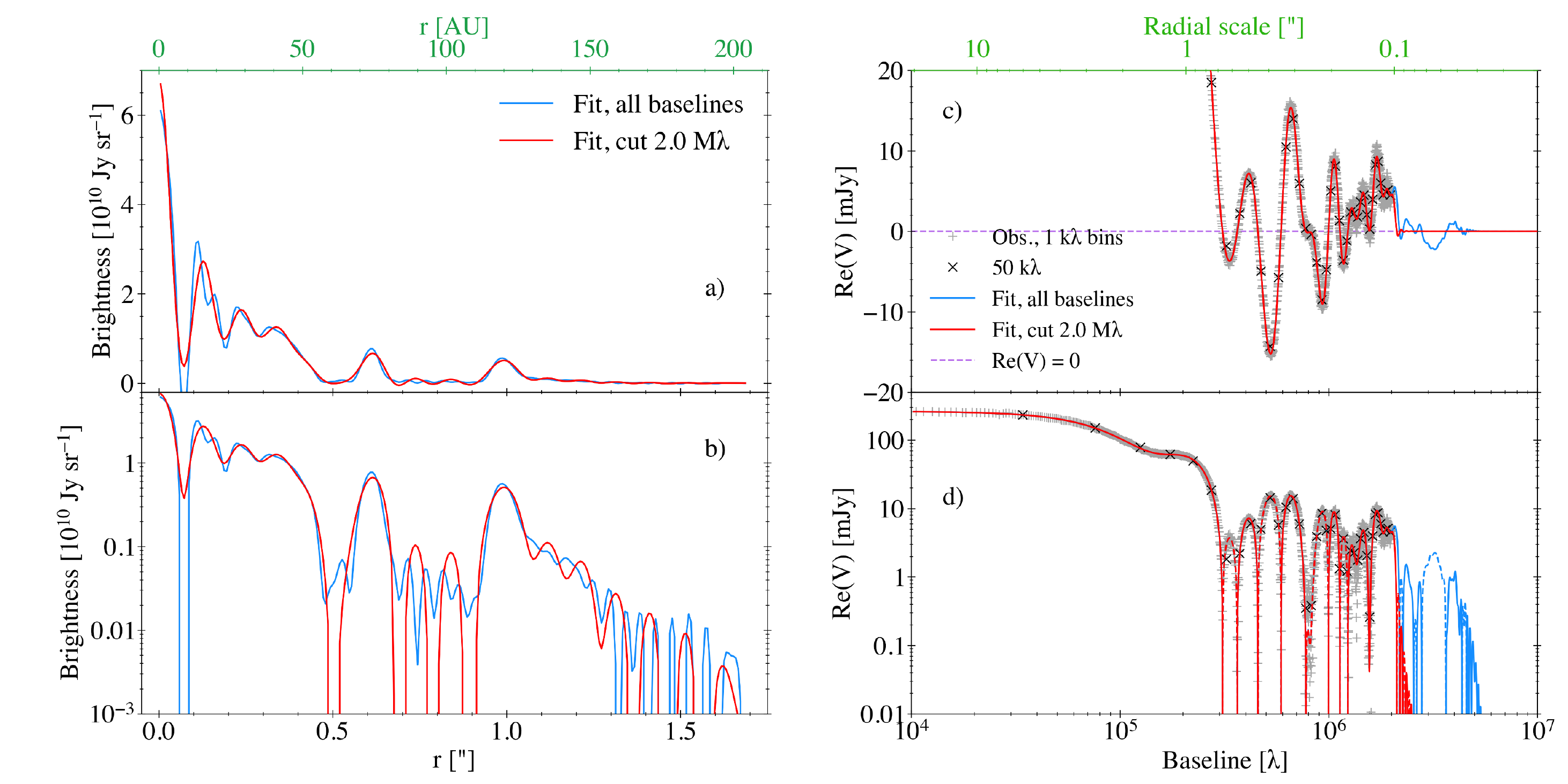}
	    \caption{{\bf Animation of AS~209 fit evolution with baseline} 
	    \newline
	    a) The fitted \fr brightness profile in Fig.~\ref{fig:as209_real}(a) and the fit's evolution as the data are truncated at successively longer maximum baseline, beginning at 1 {\ml}. \newline
	    b) As in (a) on a logarithmic y-scale. \newline
	    c) Zoom on the region around 0 of the observed DSHARP visibilities in 1 and 50 {\kl} bins. The \fr fit to the full dataset and to the data at the evolving truncation baseline are shown. \newline
	    d) As in (c) on a logarithmic y-scale. The data (and fits') negative regions are denoted by the fit lines' dashed sections. \newline
	    The animated figure is shown in the online version; a snapshot at a truncation baseline of 2 {\ml} is shown in the print version.
	    }
    \label{fig:as209_animation}
\end{figure*}

To further show that \fr can identify substructure in lower resolution observations, {\bf Fig.~\ref{fig:as209_animation}} is an animation in which the \fr fit to the DSHARP AS~209 dataset evolves as a function of the data's longest baseline. We first truncate the data at 1 {\ml} $\Leftrightarrow\ \approx$0.2$\arcsec$ prior to the visibilities being fit, then successively step the maximum baseline to 5 {\ml} and fit at each step. \fr identifies with increasing accuracy the presence and detailed morphology of substructure in the disc as the truncation baseline increases, and at each step, including when 1 {\ml} is the longest baseline, the model is correctly identifying and partially resolving more structure than a CLEAN profile obtained with the same truncated dataset. This again demonstrates that \fr can correctly identify structure in low -- moderate resolution data.

\begin{figure*}
	\includegraphics[width=\textwidth]{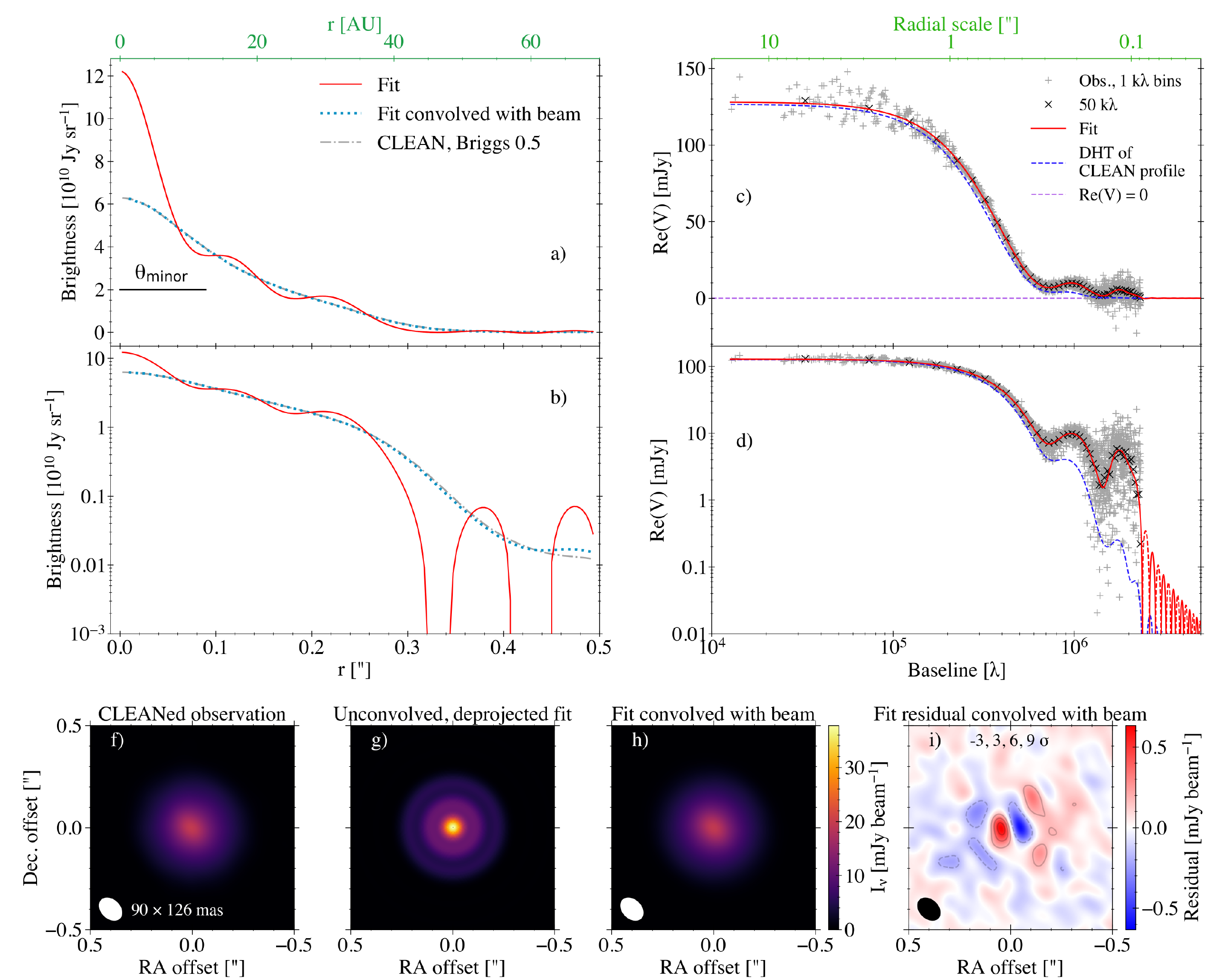}
	    \caption{ {\bf Fit to real, moderate resolution observations: DR~Tau} 
	    \newline 
	    a) Reconstructed brightness profile for the real, moderate resolution observations of DR~Tau in \citet{2019ApJ...882...49L}. A CLEAN image-extracted profile and the beam's minor axis are also shown. \newline
	    b) As in (a) on a logarithmic y-scale, emphasizing the fitted profile's outer edge at $\approx$0.3$\arcsec$ (oscillations in the \fr profile beyond this indicate the fit's noise floor). \newline
	    c) Observed visibilities in 1 and 50 {\kl} bins. The \fr fit and the discrete Hankel transform of the CLEAN profile in (a) are shown. \newline
	    d) As in (c) on a logarithmic y-scale. The \fr fit's negative regions are denoted by its dashed sections. \newline
        e) CLEANed observation. \newline
        f) Image of the \emph{unconvolved and deprojected} \fr reconstruction. \newline 
        g) Image of the \fr reconstruction reprojected and convolved with the observations' synthesized beam. \newline
        h) Residual map of the \fr reconstruction (the fit's residual visibilities imaged with CLEAN), showing evidence of non-axisymmetric structure (contour $\sigma = 60$ $\mu$Jy beam$^{-1}$, peak residual $12 \sigma$).
	    }
    \label{fig:drtau}
\end{figure*}

\fr's resolving capability for low -- moderate resolution observations also shows utility when applied to apparently smooth discs. For the FWHM $92 \times 127$ mas ($\approx$ 15 AU at 140~pc) observations of the compact ($\approx$0.3$\arcsec$) disc DR~Tau \citep{2019ApJ...882...49L}, the CLEAN image (even with uniform weighting) shows no clear signs of substructure, and the parametric visibility domain fit with a smooth profile in \citet{2019ApJ...882...49L} shows significant residual error\footnote{We used the self-calibrated measurement set in \citet{2019ApJ...882...49L}. Before extracting the visibilities using the \texttt{export\_uvtable} function of the \texttt{uvplot} package \citep{uvplot_mtazzari}, we applied channel averaging (to obtain 1 channel per spectral window) and time averaging (30 sec) to all spectral windows in the original MS table.}.
The \fr fit using default hyperparameter values in {\bf Fig.~\ref{fig:drtau}}(a) -- (b) identifies substructure and finds a factor of $\approx$2 higher peak brightness than the CLEAN profile. 
This is because the \fr fit matches the visibilities in Fig.~\ref{fig:drtau}(c) -- (d) out to the longest baselines, 2.3 {\ml} $\Leftrightarrow$ 88 mas = 12 AU. Conversely convolution with the FWHM $92 \times 127$ mas synthesized beam causes the DHT of the CLEAN profile to depart the data beyond $\approx$0.8 {\ml}.
The DHT of the CLEAN profile does track at lower accuracy the same features in the visibility distribution that \fr closely fits, and the \fr fit when convolved with the synthesized beam in Fig.~\ref{fig:drtau}(a) -- (b) matches the CLEAN profile. Together these suggest the \fr fit is correctly identifying sub-beam structure.

\fr thus infers the presence of substructure (two or more gaps) in addition to the centrally peaked component in this compact disc that appears featureless in the CLEAN image and profile. Note that \fr is still almost certainly underresolving the disc's features, identifying and localizing the presence of substructure rather than accurately characterizing the features' morphologies. 
These are governed by the longest baselines in the dataset, so while it is clear from the current data that substructure is present, its form would likely change with the addition of higher resolution data (as is the case for AS~209). We therefore caution against overinterpreting the exact form of the fitted brightness profile.

Fig.~\ref{fig:drtau}(e) -- (h) compare the CLEAN image with the image of the \emph{unconvolved and deprojected} \fr fit, the fit convolved with the beam, and the convolved residual image of the fit. The latter shows prominent asymmetric structure whose origin is not yet clear, but is not as best we can discern an artifact of an erroneous fit. 

\begin{figure}
	\includegraphics[width=\columnwidth]{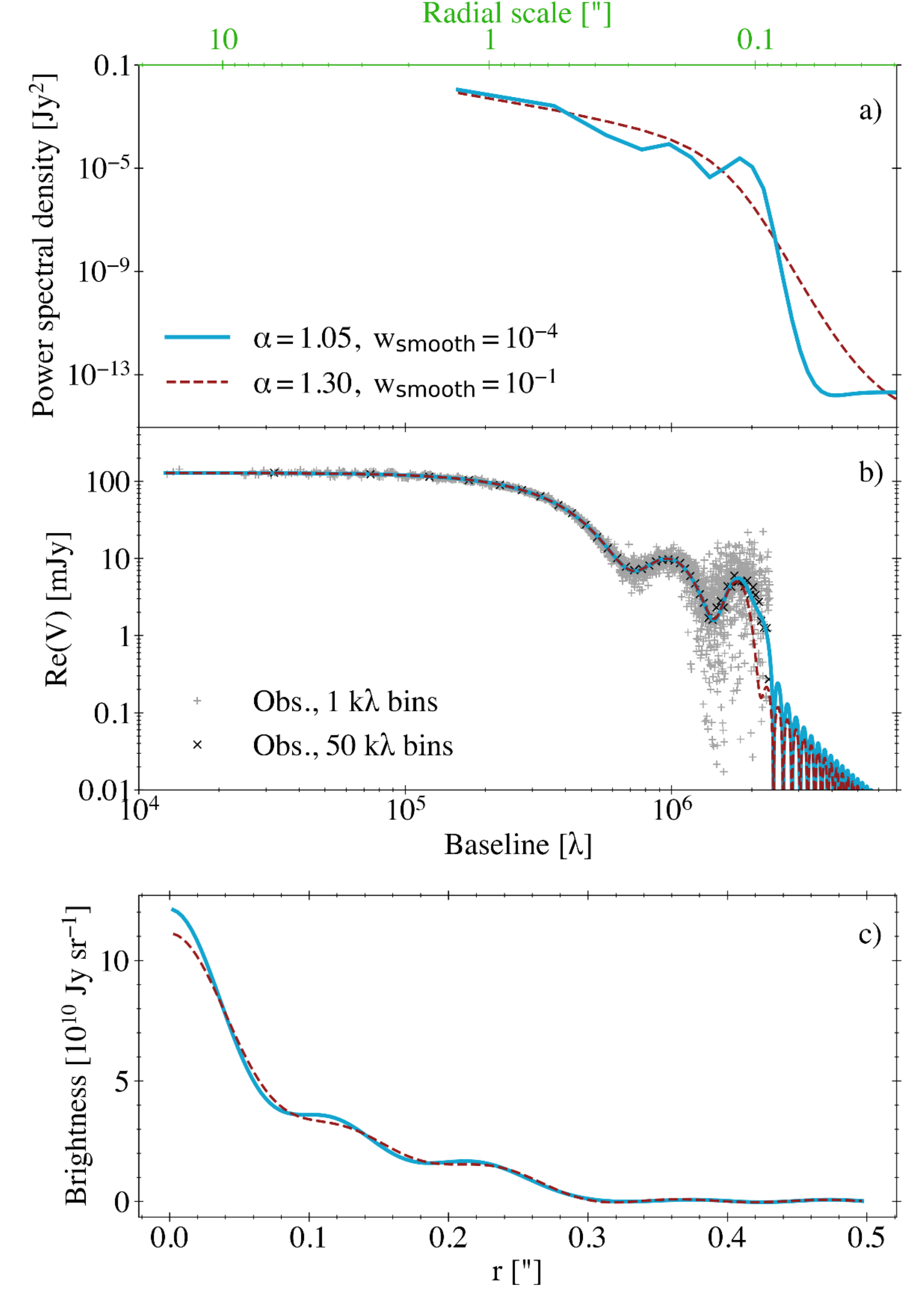}
	    \caption{{\bf Fit sensitivity to hyperparameters: DR~Tau}
	    \newline 
	    a) Maximum a posteriori power spectra of the \fr fit under the default hyperparameter values ($\alpha = 1.05,\ w_{\rm smooth} = 10^{-4}$) and stronger values ($\alpha = 1.30,\ w_{\rm smooth} = 10^{-1}$). The structural differences between these spectra show weak effect on the visibility fits in (b) and fitted profiles in (c). The power spectra are the priors placed on the respective brightness profile reconstructions. \newline
	    b) Observed visibilities in 1 and 50 k$\lambda$ bins; the \fr fit under the default hyperparameter values and under a stronger choice, showing fairly minor difference. \newline
	    c) The fitted \fr brightness profile in Fig.~\ref{fig:drtau}(a) -- (b) under the default hyperparameter values, and a fit under a stronger choice.
	    }
    \label{fig:drtau_prior}
\end{figure}

Varying the fit hyperparameters $\alpha$ and $w_{\rm smooth}$ within sensible bounds has a fairly weak effect on the \fr brightness profile as shown in {\bf Fig.~\ref{fig:drtau_prior}}. While the stronger hyperparameter choice ($\alpha = 1.30, w_{\rm smooth} = 10^{-1}$) does fit the visibilities to shorter baseline than the default hyperparameter values ($\alpha = 1.05, w_{\rm smooth} = 10^{-4}$), the effect of using the stronger values is fairly benign: slightly less prominent disc features and a peak brightness $\approx$10\% lower.

\section{Conclusions}
\label{sec:conclusions}
We have presented \fr, an open source code that uses a fast Gaussian process to recover the axisymmetric structure of sources observed with radio interferometers by directly fitting the real component of the visibilities. This work focuses on protoplanetary disc observations, though the code can be extended to applications in other physical contexts where an azimuthally averaged brightness profile is useful.

\fr's main advantages for characterizing axisymmetric structure relative to the CLEAN technique are:
\begin{itemize}
    \item {resolution}: \fr can resolve angular scales smaller than the uniform-weighted CLEAN beam while retaining sensitivity typical of a natural-weighted CLEAN image;
    \item {flexibility}: \fr yields a nonparametric reconstruction of a source's radial brightness profile, enabling it to fit the visibilities to high accuracy without additional input from the user;
    \item {speed}: \fr fits are performed in $<$1 min for datasets of $\lesssim {\rm few} \times 10^6$ visibilities;
    \item {ease of use}: \fr is a Python package with a minimum number of dependencies that is easy to install and integrate into existing codebases. 
\end{itemize}
\fr's resolution, flexibility, speed, and ease of use make it a high performance tool suitable for a wide range of applications on diverse interferometric datasets.

The major use case of \fr in the protoplanetary disc context is fitting interferometric observations to find and characterize disc substructures at higher resolution than the CLEAN algorithm. 
Recent high resolution observations of sources previously observed at lower angular resolution have begun to show the ubiquity of disc substructure, including in discs that appeared featureless at low resolution.
This motivates the utility of a technique such as \fr to resolve features and characterize discs at sub-beam scales. The model can resolve additional structure in high resolution datasets, such as the DSHARP observations of AS~209 \citep{Andrews2018}; as well as in lower resolution datasets that show featureless CLEAN images (even with uniform weighting), such as the compact disc DR~Tau \citep{Long2018}.
Specific scientific applications of the code include more accurately discerning substructure widths and amplitudes, better resolving the inner disc structure including the peak brightness, and isolating disc asymmetries in a residual image. 

\fr uses all visibilities in a dataset to inform the reconstructed brightness profile. Its resolving power is thus sensitive to variations in the baseline dependence of the data's SNR. In contrast a CLEAN image-extracted profile is largely determined by the pure baseline distribution.
Consequently for observations at any resolution, improving an observation's SNR via the on-source integration time can markedly enhance \fr's resolving capability to constrain sub-beam features, whereas it makes little difference for CLEAN.

\fr is actively developed at \frgithub, where users are welcome to contribute and to report issues.

\section*{Acknowledgements}
We are grateful to a number of community members who have freely shared their insights and ideas on the code and its applications. We especially thank the organizers and attendees of the Great Barriers Conference in Palm Cove, Australia, 21 -- 26 July 2019. Moreover we sincerely thank the referee, Ian Czekala, whose thoughtful review improved this work in multiple key areas.
JJ thanks A. Frisbie for her insights. This work was supported by the STFC consolidated grant ST/S000623/1. This work has also been supported by the European Union's Horizon 2020 research and innovation programme under the Marie Sklodowska-Curie grant agreement No. 823823 (DUSTBUSTERS). This work is part of the research programme VENI with project number 016.Veni.192.233, which is (partly) financed by the Dutch Research Council (NWO).

{\it Software:} \texttt{CASA} \citep{2007ASPC..376..127M}, \texttt{Jupyter Notebook} \citep{Kluyver:2016aa}, \texttt{Matplotlib} \citep{doi:10.1109/MCSE.2007.55}, \texttt{NumPy} \citep{doi:10.1109/MCSE.2011.37}, \texttt{SciPy} \citep{virtanen2019scipy}, \texttt{uvplot} \citep{uvplot_mtazzari}

\bibliographystyle{mnras}
\bibliography{references.bib}

\appendix

\section{Spectral smoothness hyperprior}
\label{Sec:T_matrix}
The spectral smoothness hyperprior is designed to constrain the power spectrum to be close to a power law in the absence of data. We implement this hyperprior using a numerical estimate for the integral of the second logarithmic derivative of the power spectrum,
\begin{equation}
-\frac{1}{2\sigma_{\rm s}^2} \int {\rm d} \log(q) \left(\frac{\partial^2 \log(p)}{\partial \log(q)^2}\right)^2  \approx -\frac{1}{2\sigma_{\rm s}^2} \sum_{n=2}^{N-1} \delta_{c,n} (\Delta \log{p}_n)^2, \label{eqn:T_def}
\end{equation}
as in \citet{2013PhRvE..87c2136O}. Here $\delta_{c,n} = (\log{q}_{n+1} - \log{q}_{n-1})/2$ and
\begin{equation}
    \Delta \log{p}_n = \frac{1}{\delta_{c,n}}\left(\frac{\log{p}_{n+1} - \log{p}_n}{\delta_{e,n}} -  \frac{\log{p}_{n} - \log{p}_{n-1}}{\delta_{e,n-1}}\right)
\end{equation}
is an approximation to the second logarithmic derivative of $\vec{p}$, where $\delta_{e,n} =\log{q}_{n+1} - \log{q}_n$. Equation~\ref{eqn:T_def} can be simplified as
\begin{equation}
     -\frac{1}{2\sigma_{\rm s}^2} \sum_{n=2}^{N-1} \delta_{c,n} (\Delta \log{p}_n)^2 = 
     - \frac{1}{2\sigma_{\rm s}^2} \log{\vec{p}}^T \mat{T} \log{\vec{p}},
\end{equation}
where the components of $\mat{T}$ are 
\begin{equation}
     T_{ij} = \sum_k \gamma_{ki} \delta_{c,k} \gamma_{kj},
\end{equation}
with
\begin{equation}
 \gamma_{ij} = \frac{1}{\delta_{c,i}}
  \begin{cases}
    -\left(\frac{1}{\delta_{e,i}} + \frac{1}{\delta_{e,i-1}}\right) & {\rm if } j = i \\
    +\frac{1}{\delta_{c,i\pm 1}} & {\rm if } j = i\pm1 \\
    0 & \rm{otherwise.}        
  \end{cases}
\end{equation}
The matrix $\mat{T}$ is constant and pentadiagonal, depending only on $q_k$. For large $k$, $\mat{T}$ has the asymptotic form
\begin{equation}
 T_{ij} = \frac{1}{(q_i R_{\rm max})^3} \times 
  \begin{cases}
    +48 & {\rm if } j = i \\
    -32 & {\rm if } j = i\pm1 \\
    +8 & {\rm if } j = i\pm2 \\
    0 & \rm{otherwise.}        
  \end{cases}
\end{equation}

\section{Signal-to-noise threshold}
\label{sec:SNR_thresh}
Here we show that it is the visibilities' SNR that determines whether a \fr fit ignores the data at a given baseline. This is ultimately determined by whether the maximum a posteriori value for the power spectrum parameters tends to: Equation~\ref{eq:sol_ps1}, in which case the power spectrum is determined by the visibility amplitude; or Equation~\ref{eq:sol_ps2}, in which case the power spectrum's low amplitude suppresses the power on a given scale. The argument presented here is similar to that in \citet{2011PhRvD..83j5014E}, who derive this for a general class of methods like \fr.

To make the derivation tractable, we make some simplifications: we assume $p_0 = 0$, neglect smoothing, and study the simplifying case where the visibilities are observed exactly at the spatial frequency collocation points $q_k$. While this is somewhat unrealistic, the insights derived are useful more generally. Under our last assumption, the matrix $\mat{H}(\vec{q})$ in Equation~\ref{eq:gauss_sol} is just $\mat{Y}_{\rm f}$. Now,
\begin{align}
    \mean &= \left( \mat{Y}_{\rm f}^T \mat{N}^{-1} \mat{Y}_{\rm f} + \mat{Y}_{\rm f}^T \diag{1/\vec{p}} \mat{Y}_{\rm f} \right)^{-1} \mat{Y}_{\rm f}^T \mat{N}^{-1}  \vec{V} \nonumber \\
    &= \mat{Y}_{\rm b} \left(\mat{N}^{-1} + \diag{1/\vec{p}} \right)^{-1} \mat{N}^{-1} \vec{V}.
\end{align}
This means that
\begin{equation}
    (\mat{Y}_{\rm f} \mean)_k = \frac{V_k}{1 + \sigma_k^2 / p_k}
\end{equation}
and
\begin{equation}
    (\mat{Y}_{\rm f}^T \mat{D} \mat{Y}_{\rm f})_{kk} = \frac{\sigma_k^2}{1 + \sigma_k^2 / p_k}.
\end{equation}
Using these in Equation~\ref{eq:iter} with $p_0=0$, $\sigma_{\rm s} = \infty$, and $\vec{p}^{\rm new} = \vec{p}$ (convergence to the maximum likelihood) and rearranging produces the cubic equation in $p_k$,
\begin{equation}
    p_k \left\{ (p_k + \sigma_k^2)^2 [1 + 2 (\alpha-1)] -  (p_k + \sigma_k^2) (V_k^2 + \sigma^2) - V_k^2 \sigma^2\right\} = 0,
\end{equation}
which always has the solution $p_k = 0$. Solutions for $p_k > 0$ can be found by completing the square of the term in brackets;
\begin{align}
    p_k = &\frac{1}{1 + 2(\alpha-1)} \times \nonumber \\
    &\frac{1}{2}\bigg\{\left[V_k^2 - \sigma^2(1+4(\alpha-1))\right] + \nonumber \\
                        & \qquad \qquad \qquad \qquad \sqrt{(V_k^2 - \sigma_k^2)^2 - 8V_k^2\sigma_k^2(\alpha-1)} \bigg\} . \label{eq:ps_append}
\end{align}
For $V_k \gg \sigma_k$, this yields an equivalent expression to that given in Equation~\ref{eq:sol_ps1}. However, the term inside the square root in Equation~\ref{eq:ps_append} is only positive if
\begin{equation}
    V_k^2 > \sigma^2 \left\{1 + 4(\alpha-1) + 2\sqrt{2(\alpha-1)[1 + 2(\alpha-1)]}\right\},
\end{equation}
otherwise $p_k=0$ is the only solution (note that in the special case $\alpha=1$, the term in brackets in Equation~\ref{eq:ps_append} is always positive; however the above expression still correctly denotes the region for which solutions with $p_k > 0$ exist). The implication of this is that for an SNR below a given threshold, $V_k^2 \lesssim \sigma_k^2$, the inverse $\Gamma$ hyperprior will drive the power spectrum toward zero. In practice, including a nonzero $p_0$ means that $p_k \rightarrow p_0 / (\alpha-1)$, as given in Equation~\ref{eq:sol_ps2}.

Equation~\ref{eq:ps_append} also shows why we use $\alpha>1$. At long baselines -- where the visibilities are noise-dominated -- we will have $V_k^2 \sim \sigma_k^2$, and there will be many statistical fluctuations causing $V_k$ to be slightly greater than $\sigma_k$. Since for $\alpha=1$ the threshold is exactly at $V_k = \sigma_k$, the power spectrum will contain some fraction of this power, and the model will fit the noise-dominated visibilities. Increasing $\alpha$ to 1.05, 1.10, 1.30, or 1.90 increases the noise threshold to 1.36, 1.54, 2.04, or 3.01 $\sigma_k$ respectively. 
Thus using $\alpha > 1$ markedly reduces the chance that the model will attempt to fit noise-dominated data.

A similar argument can be made in the case where the visibilities are not observed directly at the spatial frequency collocation points, so long as the ($u,v$) plane is sampled well enough that $\mat{M}$ can be approximated as a diagonal matrix in the visibility domain. In this case $V_k = \sigma_k^2 (\mat{Y}_{\rm b}^T \vec{j})_k$, where $\sigma_k^2 \approx (\mat{Y}_{\rm f}^T \mat{M} \mat{Y}_{\rm f})_{kk}^{-1}$ is the effective variance at that scale.

\section{Nonnegative fits}
\label{sec:non_negative}
In some circumstances it may be beneficial to have brightness profiles that are nonnegative, i.e., $\vec{I}_\nu \ge 0$, as is produced by maximum entropy methods or a log-normal model for the brightness \citep{2016A&A...586A..76J}. In \fr we can generate nonnegative solutions by finding the most probable brightness reconstruction $\vec{I}_\nu$ for a given set of power spectrum parameters and the constraint $\vec{I}_\nu \ge 0$ (i.e., the maximum of $P(\vec{I}_\nu | \vec{p})$ subject to $\vec{I}_\nu \ge 0$) using the nonnegative least squares solver in the \textsc{scipy} package (\texttt{scipy.optimize.nnls}).

\begin{figure}
	\includegraphics[width=\columnwidth]{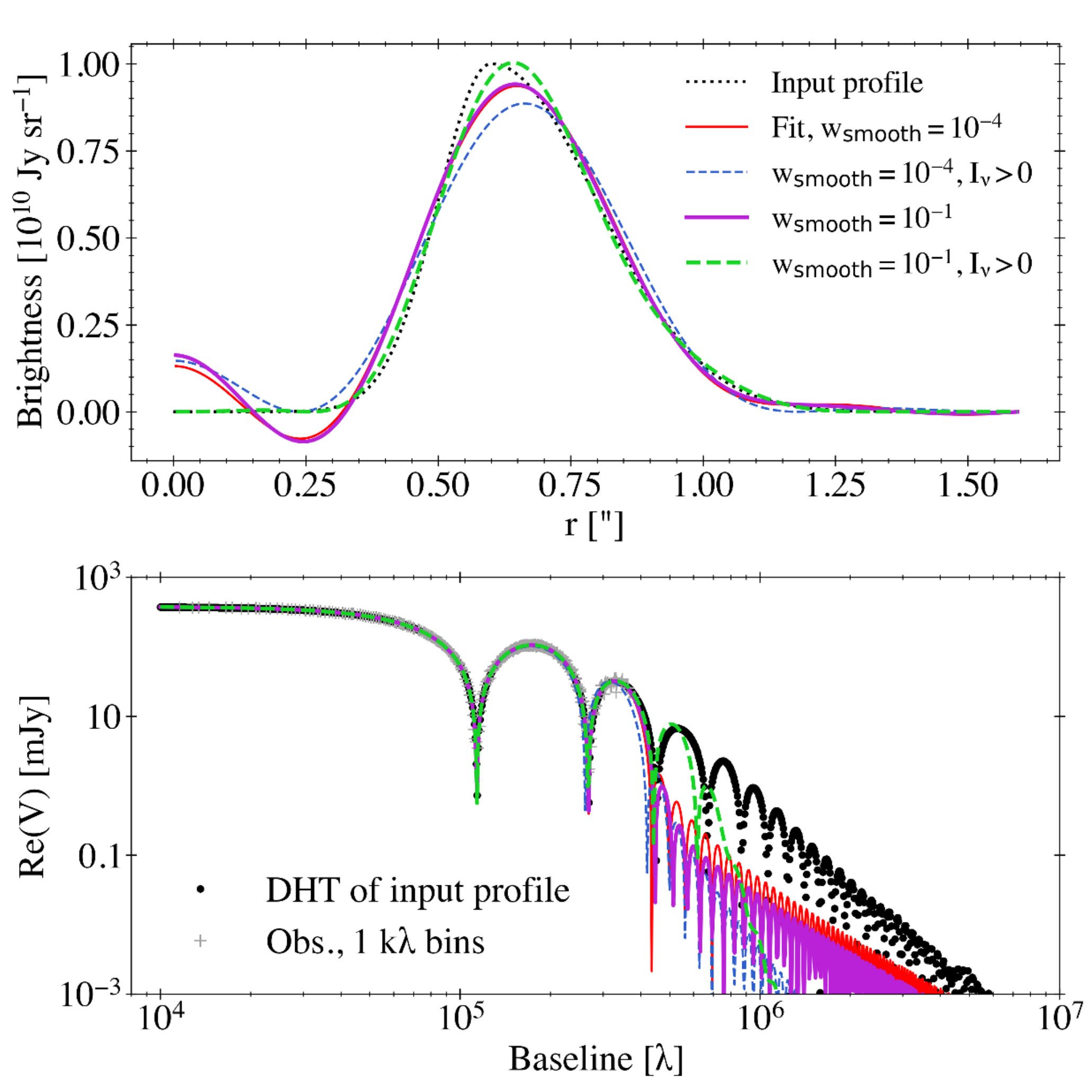}
	    \caption{{\bf Comparison of standard and nonnegative fits}
	    \newline
	    a) Input and reconstructed brightness profiles for a mock Gaussian ring (two joined sigmoids) observed with the ALMA C43-3 configuration (synthesized beam FWHM $0.59 \times 0.70 \arcsec$, Briggs=0.5; see Table~\ref{tab:obs} and Sec.~\ref{sec:convolution}), shown under two values of the $w_{\rm smooth}$ hyperparameter, and with brightness profile positivity either unenforced or enforced. All fits use $\alpha = 1.05$. \newline
	    b) Observed visibilities in 1 k$\lambda$ bins, the DHT of the noiseless input profile, and the \fr fits corresponding to the profiles in (a).
        }
    \label{fig:non_negative}
\end{figure}

{\bf In Fig.~\ref{fig:non_negative}} we show the impact of the $\vec{I}_\nu \ge  0$ constraint on the reconstructed brightness profile for the mock Gaussian ring under the C43-3 mock observation.
Comparing the posterior mean brightness profile and the most probable nonnegative solution under two values of $w_{\rm smooth}$, Fig.~\ref{fig:non_negative}(a) shows the nonnegative solution is more accurate. Correspondingly the constraint $\vec{I}_\nu \ge 0$ yields a visibility fit in Fig.~\ref{fig:non_negative}(b) that initially falls slower than the standard \fr fit beyond the data's longest baseline (0.4 {\ml}), predicting the visibility amplitude at unobserved baselines more accurately than the standard fit between 0.4 -- 0.8 {\ml}.

The improvement gained by enforcing $\vec{I}_\nu \ge  0$ does however depend on the hyperparameters, with the nonnegative fit using $w_{\rm smooth} = 10^{-4}$ fitting the visibilities less accurately than the analogous standard \fr fit. This hyperparameter dependency is a result of the need to introduce additional power on small scales when enforcing a nonnegative brightness.
Large $\alpha$ and small $w_{\rm smooth}$ strongly damp power at long baseline (on small spatial scales), introducing tension between the prior and the $\vec{I}_\nu \ge 0$ constraint. 
This can cause the nonnegative reconstruction to be a poor fit to the data on shorter baselines. Conversely with smaller $\alpha$ and larger $w_{\rm smooth}$, the maximum a posteriori power spectra fall off more slowly with increasing $q$, and the resulting nonnegative brightness reconstruction has more power on small spatial scales.

Because this process involves extrapolating the visibilities beyond their maximum baseline (or at least into a region where they are noise-dominated), we therefore urge caution when interpreting any additional structures introduced by enforcing the profile be nonnegative.

\section{Mock disc functional forms}
Table~\ref{tab:functional_forms} gives the functional forms of the brightness profiles for the mock discs shown in Fig.~\ref{fig:morphologies} and listed in Table~\ref{tab:obs}.
\begin{table*}
\caption{Functional forms of the brightness profiles for the archetypal mock discs listed in Table~\ref{tab:obs} and shown in Fig.~\ref{fig:morphologies}.}
\begin{tabular}{l l c c}
    \hline
    Disc & Brightness profile functional form & Amplitude variables & Location variables [$\arcsec$] \\
    & & [$10^{10}$ Jy sr$^{-1}$] & \\
    \hline
    Gaussian centered at 0 & $I_\nu(r) = a e^{\frac{-r^2}{2 c^2}}$ & $a = 1.0$ & $c = 0.2$ \\ 
    \\
    Gaussian ring & 
    \[ 
     I(r) = 
     \begin{cases} 
      a e^{\frac{-(r-b)^2}{2 c_1^2}}, & r \leq b  \\
      a e^{\frac{-(r-b)^2}{2 c_2^2}}, & r > b \\
     \end{cases}
    \]
    & $a = 1.0$ & \thead{$b = 0.6$ \\ $c_1 = 0.1$ \\ $c_2 = 0.2$} \\
    \\
    Multi-Gaussian & 
    \[ 
     I(r) = a_1 e^{\frac{-(r - b_1)^2}{2 c_1^2}} - a_2 e^{\frac{-(r - b_2)^2}{2 c_2^2}} - a_3 e^{\frac{-(r - b_3)^2}{2 c_3^2}} +
     \begin{cases} 
      a_4 e^{\frac{-(r-b_4)^2}{2 c_4^2}}, & r \leq b_4  \\
      a_4 e^{\frac{-(r-b_4)^2}{2 c_5^2}}, & r > b_4 \\
     \end{cases}
    \] 
    & \thead{$a_1 = 0.6$ \\ $a_2 = 0.4$ \\ $a_3 = 0.3$ \\ $a_4 = 1.0$} & \thead{$b_1 = 0.70$ \\ $b_2 = 0.60$ \\ $b_3 = 0.85$ \\ $b_4 = 0.15$ \\ $c_1 = 0.20$ \\ $c_2 = 0.05$ \\ $c_3 = 0.05$ \\ $c_4 = 0.05$ \\ $c_5 = 0.10$} \\
    \\
    Sharp-edged & 
    \[ 
     I(r) = e^{-2r^2} \cdot \left(a_1 + 
     \begin{cases} 
      a_2, & b_1 \leq r \leq b_2 \\
      a_3, & b_3 \leq r \leq b_4 \\
      a_4, & b_5 \leq r \leq b_6 \\
      a_5, & b_7 \leq r \leq b_8 \\
      0, & {\rm else} \\
     \end{cases}
    \right)
    \] 
    & \thead{$a_1 = 1.00$ \\ $a_2 = -0.90$ \\ $a_3 = -0.20$ \\ $a_4 = -0.25$ \\ $a_5 = 0.50$}  & \thead{$b_1 = 0.1$ \\ $b_2 = 0.2$ \\ $b_3 = 0.4$ \\ $b_4 = 0.7$ \\ $b_5 = 0.6$ \\ $b_6 = 0.7$ \\ $b_7 = 0.7$ \\ $b_8 = 0.8$} \\
    \hline
\end{tabular}
\label{tab:functional_forms} 
\end{table*}

\bsp
\label{lastpage}
\end{document}